\documentclass[preprint]{aastex}
\usepackage{amsmath}
\usepackage{float}
\usepackage{multirow} 

\def\Rsun{R$_{\odot}$}
\def\Msun{M$_{\odot}$}
\def\Zsun{Z$_{\odot}$}
\newcommand{\appropto}{\mathrel{\vcenter{
  \offinterlineskip\halign{\hfil$##$\cr
    \propto\cr\noalign{\kern2pt}\sim\cr\noalign{\kern-2pt}}}}}

\begin{document}

\title{The Close Binary Properties of Massive Stars \\
in the Milky Way and Low-Metallicity Magellanic Clouds}

\author{Maxwell Moe\altaffilmark{1} \& Rosanne Di Stefano\altaffilmark{1}}

\altaffiltext{1}{Harvard-Smithsonian Center for Astrophysics, 60 Garden Street, MS-10, Cambridge, MA, 02138, USA; mmoe@cfa.harvard.edu}

\begin{abstract}

 In order to understand the rates and properties of Type Ia and Type Ib/c supernovae, X-ray binaries, gravitational wave sources, and gamma ray bursts as a function of galactic environment and cosmic age, it is imperative that we measure how the close binary properties of O and B-type stars vary with metallicity.  We have studied eclipsing binaries with early-B main-sequence primaries in three galaxies with different metallicities: the Large and Small Magellanic Clouds (LMC and SMC, respectively) as well as the Milky Way (MW).  The observed fractions of early-B stars which exhibit deep eclipses 0.25~$<$~$\Delta$m\,(mag)~$<$~0.65 and orbital periods 2~$<$~$P$\,(days)~$<$~20 in the MW, LMC, and SMC span a narrow range of (0.7\,-\,1.0)\%, which is a model independent result.  After correcting for geometrical selection effects and incompleteness toward low-mass companions, we find for early-B stars in all three environments: (1) a close binary fraction of (22\,$\pm$\,5)\% across orbital periods 2~$<$~$P$\,(days)~$<$~20 and mass ratios $q$~=~$M_2$/$M_1$~$>$~0.1, (2) an intrinsic orbital period distribution slightly skewed toward shorter periods relative to a distribution that is uniform in log\,$P$, (3) a mass-ratio distribution weighted toward low-mass companions, and (4) a small, nearly negligible excess fraction of twins with $q$~$>$~0.9. Our fitted parameters derived for the MW eclipsing binaries match the properties inferred from nearby, early-type spectroscopic binaries, which further validates our results.  There are no statistically significant trends with metallicity, demonstrating that the close binary properties of massive stars do not vary across metallicities $-$0.7~$<$~log($Z$/\Zsun)~$<$~0.0 beyond the measured uncertainties.  
\end{abstract}


\section{Introduction}

Spectral type O ($M_1$ $\gtrsim$ 18\,\Msun) and B (3\,\Msun\ $\lesssim$ $M_1$ $\lesssim$ 18\,\Msun)  primaries with close binary companions evolve to produce a plethora of astrophysical phenomena, including millisecond pulsars \citep{Lorimer2008}, Type Ia \citep{Wang2012} and possibly Type Ib/c \citep{Yoon2010} supernovae, X-ray binaries \citep{Verbunt1993}, Algols \citep{vanRensbergen2011}, short \citep{Nakar2007} and perhaps long \citep{Izzard2004} gamma ray bursts, accretion induced collapse \citep{Ivanova2004}, and gravitational waves \citep{Schneider2001}.   Telescopic surveys dedicated to discovering luminous transients and/or high-energy sources have identified some of these binary star phenomena in low-metallicity host environments such as dwarf and high-redshift galaxies \citep{Kuznetsova2008,McGowan2008,Berger2009,Frederiksen2012}.  Recent observations have demonstrated that the rates and properties of certain channels of binary evolution vary with metallicity \citep{Dray2006,Cooper2009,Sullivan2010,Kim2013}. To explain these observed trends, it has been postulated that the physical processes that affect stellar and binary evolution are metallicity dependent \citep{Bellazzini1995,Kobayashi1998,Ivanova2006,Fryer2007,Kistler2011}. However, the initial conditions of the progenitor main-sequence (MS) binaries may change with metallicity \citep{Machida2008}, which may also account for the observations.  In order to distinguish between these two hypotheses, it is imperative that we measure the close binary properties of massive stars at low metallicity.

In the MW, the fraction of primaries which harbor close companions dramatically increases with primary mass \citep[][see also \S4]{Abt1983,Raghavan2010}, reaching $\approx$70\% with orbital periods $P$ $<$ 3,000 days for massive O-type stars \citep{Sana2012}.  Yet the effect of metallicity on the close binary fraction of massive stars has not been robustly measured from observations.  This is primarily due to the paucity of short-lived, low-metallicity early-type stars within our own Milky Way (MW), forcing us to explore external galaxies to investigate metallicity dependence.  \citet{Evans2006} utilized multi-epoch spectroscopic observations of massive stars in the Large and Small Magellanic Clouds (LMC and SMC, respectively) to derive a lower limit of $\approx$30\% for the close binary fraction.  Their cadence was insufficient to fit orbital periods to their radial velocity data for many of their systems, so they were unable to account for incompleteness. \citet{Sana2013} searched for spectroscopic binaries among O-type stars in the starburst region of the Tarantula Nebula, also known as 30 Doradus, within the LMC.  After correcting for observational biases, they computed a binary fraction of $\approx$50\% across orbital periods 0.15 $<$ log\,$P$\,(days) $<$ 3.5.  This extremely active and dense environment may not be representative of all O-type stars.  Moreover, with slightly subsolar abundances of [Fe/H] $\approx$ [O/H] $\approx$ $-$0.2 \citep{Peimbert2010}, 30~Doradus offers little leverage to gauge the effect of metallicity.   Finally, \citet{Mazeh2006} utilized observations made during the second phase of the Optical Gravitational Lensing Experiment (OGLE-II) to identify eclipsing binaries with B-type primaries in the LMC.  After correcting for geometrical and other selection effects, they estimated that only $\approx$0.7\% of B stars have a companion with orbital periods $P$~=~2~$\mbox{-}$~10~days, nearly an order of magnitude lower than the value for Milky Way counterparts inferred from spectroscopic radial velocity observations.  However, \citet{Mazeh2006} did not account for incompleteness towards low mass secondaries, so it is conceivable that many small companions are hiding by exhibiting shallow eclipses below the threshold of the OGLE-II sensitivity.  

In this paper, we analyze catalogs of eclipsing binaries in the MW, LMC, and SMC to determine the close binary fraction of early-B stars as a function of metallicity.  We organize the subsequent sections as follows.  In \S2, we discuss the criteria we developed to compile our samples of eclipsing binaries from various catalogs, and compare the observed properties of the eclipsing systems among the different environments. In \S3, we utilize sophisticated light curve modeling software and perform detailed Monte Carlo simulations to correct for observational selection effects and incompleteness. In \S4, we compare our results derived from eclipsing binaries to spectroscopic radial velocity observations of O and B-type binaries in the MW. We summarize and discuss our conclusions in \S5.

\section{The Eclipsing Binary Samples}

  We utilize catalogs of eclipsing binaries in the MW based on Hipparcos data \citep{Lefevre2009}, in the LMC identified by OGLE-II \citep{Wyrzykowski2003} and OGLE-III observations \citep{Graczyk2011}, and in the SMC discovered by the OGLE-II survey \citep{Wyrzykowski2004}.  These surveys identified eclipsing systems with varying sensitivity and completeness.  In order to make accurate comparisons among these catalogs, we must first apply selection criteria to create a uniform dataset.  

First, we select relatively unevolved $M_1$ $\approx$ 7\,\Msun\ - 18\,\Msun\ primaries, corresponding to spectral types $\approx$B0-B3.5 and luminosity classes $\approx$III-V.    By selecting a narrow range of spectral types and stages of evolution, we can more robustly correct for geometrical selection effects and other observational biases (see \S3).  Because the mass function of early-B stars is strongly skewed toward lower mass objects, the median primary mass in our selected samples is $M_1$ = 10\,\Msun\ (see \S3.1).  

Second, we restrict our samples to eclipsing binaries with orbital periods $P$ = 2 - 20 days. We do not consider shorter period binaries with $P$ $<$ 2 days because a large fraction of these systems are contact binaries (EW eclipsing types / W Ursae Majoris variables) that may have substantially evolved from their primordial configurations.  Eclipsing binary identification algorithms typically fail to detect MS binaries when the eclipse duration is $\lesssim$5\% the total orbital period \citep{Soderhjelm2005}.  For our early-B primaries with MS companions, the eclipse widths fall below 4\% the total orbital period when the orbital period exceeds $P$ = 20 days (see \S3.1). 

Finally, we select eclipsing binaries within a particular range of primary eclipse depths $\Delta$m.  For spherical MS stars, the maximum eclipse depth possible is $\Delta$m = 0.75 mag, corresponding to a twin system with equal mass components observed edge-on at inclination $i$ = 90$^{\rm o}$. In a real stellar population, eclipsing binaries with $\Delta$m $\gtrsim$ 0.65 are significantly contaminated by systems which have undergone binary evolution, e.g. Algols \citep[][see their Figure 5]{Soderhjelm2005}, and/or are substantially tidally distorted, so we only consider systems with $\Delta$m $<$ 0.65. Because we selected eclipsing binaries with relatively unevolved primaries and $P$ $>$ 2 days, most systems with $\Delta$m~$<$~0.65 in our samples are not filling their Roche lobes (see also \S3.1).  Depending on the photometric accuracy, the catalogs become less sensitive toward shallow eclipse depths $\Delta$m~$\lesssim$~0.10$\,\mbox{-}\,$0.25.  We consider two subsamples: deep eclipses with 0.25 $<$ $\Delta$m $<$ 0.65 where all the surveys are sensitive, and an extension that also includes medium eclipse depths with 0.10~$<$~$\Delta$m~$<$~0.65 where only some of the samples are still complete.  

Nearby early-B stars in the MW within $\approx$2 kpc of our sun cover a narrow range of metallicities centered on solar composition (\citealt{Gummersbach1998}, [O/H]~=~$-$0.2\,$\pm$\,0.2, [Mg/H]~=~0.0\,$\pm$\,0.2; \citealt{Daflon2004}, [O/H]~=~$-$0.1\,$\pm$\,0.2, [Mg/H]~=~$-$0.1\,$\pm$\,0.2; \citealt{Lyubimkov2005}, [Mg/H]~= 0.1\,$\pm$\,0.2). Although most catalogs of eclipsing binaries in the MW focus on lower mass, solar-type primaries, \citet{Lefevre2009} recently classified a list of variable O and early-B stars based on Hipparcos data.  They identified ${\cal N}_{\rm EB}$ = 51 eclipsing binaries with $P$ = 2\,-\,20 days, median Hipparcos magnitudes $\langle$H$_{\rm P}\rangle$~$<$~9.3, and primaries displaying either spectral types B0-B2 and luminosity classes III-V or spectral types B2.5-B3 and luminosity classes II-V.  From these systems, ${\cal N}_{\rm med}$~=~31 exhibited eclipse depths 0.10~$<$~$\Delta$H$_{\rm P}$~$<$~0.65, while only ${\cal N}_{\rm deep}$ = 16 had deep amplitudes 0.25~$<$~$\Delta$H$_{\rm P}$~$<$~0.65.  In the Hipparcos database \citep{Perryman1997}, there are ${\cal N}_{\rm B}$~=~1596 early-B stars which satisfy the same magnitude, spectral type, and luminosity class criteria, where we have included objects without a specifically listed luminosity class but excluded B0-B2 spectral types with a hybrid II-III designation.  This results in ${\cal F}_{\rm  med}$~=~${\cal N}_{\rm med}/{\cal N}_{\rm B}$~=~(1.94\,$\pm$\,0.35)\% and ${\cal F}_{\rm deep}$~=~${\cal N}_{\rm deep}/{\cal N}_{\rm B}$~=~(1.00\,$\pm$\,0.25)\%, where the errors derive from Poisson statistics.\footnote{Throughout this work, we use ${\cal N}$ to represent an absolute {\it number},  ${\cal F}$ for a {\it fraction}, either observed or intrinsic, ${\cal O}$ to represent an {\it observed} distribution which integrates to the specified fraction, ${\cal S}$ for a {\it simple} approximation to the observed distribution, ${\cal M}$ for a detailed {\it model} distribution based on our Monte Carlo simulations, ${\cal U}$ for an intrinsic distribution which describes the {\it underlying} close binary population, ${\cal C}$ for a {\it correction} factor, ${\cal P}$ for the {\it probability} that a close binary is observed as an eclipsing system,  and $p$ for either a {\it probability} density distribution which integrates to unity or a {\it probability} statistic from a hypothesis test.}  We summarize these results in Table 1.

\begin{table}[H]
\fontsize{10}{15}
\selectfont
\setlength{\tabcolsep}{3pt}
\center
\begin{tabular}{|c|r|c|r|r|r|c|r|c|c|}
\hline
Galaxy  & $\langle$log(Z/\Zsun)$\rangle$ & Survey & ${\cal N}_{\rm B}$~~ & ${\cal N}_{\rm EB}$~  & ${\cal N}_{\rm med}$~ & ${\cal F}_{\rm med}$ & ${\cal N}_{\rm deep}$ & ${\cal F}_{\rm deep}$ & Refs \\
\hline
MW  &    0.0~~~~~ & Hipparcos &  1,596~ &    51~ &     \bf{31}~ & \bf{(1.94$\pm$0.35)\%} &  \bf{16}~ & \bf{(1.00$\pm$0.25)\%} & 1,2 \\
\hline
LMC & $-$0.4~~~~~ &  OGLE-II  & 20,974~ &   308~ &         263~ & (1.25$\pm$0.08)\%      & \bf{145}~ & \bf{(0.69$\pm$0.06)\%} & 3,4 \\
\hline
LMC & $-$0.4~~~~~ &  OGLE-III & 69,616~ & 2,024~ &  \bf{1,301}~ & \bf{(1.87$\pm$0.05)\%} & \bf{477}~ & \bf{(0.69$\pm$0.03)\%} & 5,6 \\
\hline
SMC & $-$0.7~~~~~ &  OGLE-II  & 21,035~ &   298~ &         277~ & (1.32$\pm$0.08)\%      & \bf{147}~ & \bf{(0.70$\pm$0.06)\%} & 7,8 \\
\hline
\end{tabular}
\caption{Eclipsing binary statistics of early-B MS stars in the Milky Way and Magellanic Clouds.  The first three columns give the host galaxy, mean metallicity of early-type stars (see text for details), and survey from which the eclipsing binaries were identified.  Column 4 lists the total number ${\cal N}_{\rm B}$ of relatively unevolved early-B primaries in the samples, while column 5 gives the number ${\cal N}_{\rm EB}$ of eclipsing binaries with orbital periods $P$ = 2\,-\,20 days. Columns 6 and 7 list the numbers ${\cal N}_{\rm med}$ and fractions ${\cal F}_{\rm med}$ = ${\cal N}_{\rm med}$/${\cal N}_{\rm B}$  of systems with eclipse depths $\Delta$m~=~0.10\,-\,0.65~mag and orbital periods $P$ = 2\,-\,20 days.  Columns 8 and 9 give similar numbers ${\cal N}_{\rm deep}$ and fractions ${\cal F}_{\rm deep}$~=~${\cal N}_{\rm deep}$/${\cal N}_{\rm B}$, but for those systems displaying deep eclipses $\Delta$m~=~0.25\,-\,0.65~mag only. Shown in boldface are the cases for which the samples are relatively complete, i.e. when the photometric accuracy of the survey is sensitive to the specified eclipse depths.   1 - \citet{Perryman1997}; 2 - \citet{Lefevre2009}; 3 - \citet{Udalski2000}; 4 - \citet{Wyrzykowski2003}; 5 - \citet{Udalski2008}; 6 - \citet{Graczyk2011}; 7 - \citet{Udalski1998}; 8 - \citet{Wyrzykowski2004}.}
\end{table}

The LMC provides our first testbed to investigate the effects of metallicity on the frequency of close early-B binaries.  Young massive stars and Cepheids, which recently evolved from B-type MS progenitors, have a mean metallicity of $\langle$log(Z/\Zsun)$\rangle$ = $-$0.4 in this nearby satellite galaxy (\citealt{Luck1998}, [Fe/H]~=~$-$0.3\,$\pm$\,0.2;  \citealt{Korn2000}, [Fe/H]~$\approx$~$-$0.4; \citealt{Rolleston2002}, [O/H]~=~$-$0.3\,$\pm$\,0.1, [Mg/H]~=~$-$0.5\,$\pm$\,0.2; \citealt{Romaniello2005}, [Fe/H]~=~$-$0.4\,$\pm$\,0.2; \citealt{Keller2006}, [Fe/H]~=~$-$0.3\,$\pm$\,0.2), where \Zsun\ = 0.015 \citep{Lodders2003,Asplund2009}. The LMC has a distance modulus of $\mu$ = 18.5, typical reddening of E(V$-$I)~=~0.1, and average extinction of A$_{\rm V}$ = 0.4 toward younger stellar environments \citep{Zaritsky1999,Imara2007,Haschke2011,WagnerKaiser2013}.   We therefore use M$_{\rm I}$~=~m$_{\rm I}$~$-$~18.8 to convert apparent magnitudes to intrinsic absolute magnitudes for the LMC.  We select relatively unevolved early-B stars with observed colors V$-$I $<$ 0.1 and  absolute magnitudes $-$3.8~$<$~M$_{\rm I}$~$<$~$-$1.5 \citep[][see also \S3.1.1]{Cox2000,Bertelli2009}. 

For the LMC, we compare the regularly monitored OGLE-II fields, which covered 4.6 square degrees in the central portions of the galaxy, to the recent OGLE-III data, which extended an additional 35 square degrees into the periphery.  We expect these two populations to be similar since there is no significant metallicity gradient in the LMC \citep{Grocholski2006,Piatti2013}.  In the central fields of the OGLE-II LMC photometric catalog \citep{Udalski2000}, ${\cal N}_{\rm B}$~=~20,974 stars have 15.0 $<$ I $<$ 17.3 and V$-$I $<$ 0.1. \citet{Wyrzykowski2003} utilized an automated search algorithm to discover eclipsing binaries in the OGLE-II LMC data, and found ${\cal N}_{\rm EB}$~=~308 systems which meet our magnitude and color cuts as well as have orbital periods between 2 and 20 days. Of these systems, ${\cal N}_{\rm med}$~=~263 have primary eclipse depths 0.10~$<$~$\Delta$I~$<$~0.65, resulting in ${\cal F}_{\rm med}$ = (1.25\,$\pm$\,0.08)\%, while ${\cal N}_{\rm deep}$~=~145 have 0.25 $<$ $\Delta$I $<$ 0.65, giving  ${\cal F}_{\rm deep}$~=~(0.69\,$\pm$\,0.06)\%.  In the larger OGLE-III LMC footprint of 35 million objects \citep{Udalski2008}, ${\cal N}_{\rm B}$~=~69,616 stars remain after we apply the same magnitude and color cuts.  \citet{Graczyk2011} used these observations to identify eclipsing binaries, being careful to exclude non-eclipsing phenomena such as ellipsoidal variables, pulsators, etc. They found ${\cal N}_{\rm EB}$~=~2,024 eclipsing binaries with primary eclipse periods $P$ = 2\,-\,20 days and photometric properties which satisfy our selection criteria. From these eclipsing binaries, ${\cal N}_{\rm med}$~=~1,301 have 0.10 $<$ $\Delta$I $<$ 0.65 and ${\cal N}_{\rm deep}$~=~477 have 0.25~$<$~$\Delta$I~$<$~0.65, giving ${\cal F}_{\rm med}$~=~(1.87\,$\pm$\,0.05)\% and  ${\cal F}_{\rm deep}$~=~(0.69\,$\pm$\,0.03)\%, respectively.  We display these LMC results for both the OGLE-II and OGLE-III samples in Table 1.

Young B stars and massive Cepheids in the SMC exhibit even lower metallicities of $\langle$log(Z/\Zsun)$\rangle$~= $-$0.7 (\citealt{Luck1998}, [Fe/H]~=~$-$0.7\,$\pm$\,0.1; \citealt{Korn2000}, [Fe/H]~$\approx$~$-$0.7; \citealt{Romaniello2005}, [Fe/H]~=~$-$0.7\,$\pm$\,0.1; \citealt{Keller2006}, [Fe/H]~=~$-$0.6\,$\pm$\,0.1), providing even greater leverage to test the effects of metallicities. Compared to the LMC, the SMC is farther away with  $\mu$~=~19.0, and experiences similar reddening and extinction of E(V$-$I) = 0.1 and A$_{\rm V}$ = 0.4 \citep{Zaritsky2002,Haschke2012}.  We therefore use M$_{\rm I}$ = m$_{\rm I}$ $-$ 19.3 and apply the same color and absolute magnitude cuts that we implemented above for the LMC.    There are ${\cal N}_{\rm B}$~=~21,035 stars with 15.5~$<$~I~$<$~17.8 and V$-$I $<$ 0.1 in the 2.4 square degree OGLE-II SMC field \citep{Udalski1998}.  From these primaries, \citet{Wyrzykowski2004} found ${\cal N}_{\rm EB}$~=~298 eclipsing binaries with $P$~=~2\,-\,20 days.  A total of ${\cal N}_{\rm med}$~=~277 of these systems have 0.10 $<$ $\Delta$I $<$ 0.65, giving ${\cal F}_{\rm med}$ = (1.32\,$\pm$\,0.08)\%, and ${\cal N}_{\rm deep}$~=~147 have 0.25~$<$~$\Delta$I~$<$~0.65, resulting in  ${\cal F}_{\rm deep}$ = (0.70\,$\pm$\,0.06)\%.  We tabulate these SMC results in Table~1.

We first compare the deep eclipsing binary fractions ${\cal F}_{\rm deep}$ of the different populations listed in Table 1.  All four surveys were sensitive to these deep eclipses, so that ${\cal F}_{\rm deep}$ should be complete.  Remarkably, the three OGLE Magellanic Cloud values match each other within the observational uncertainty of $\approx$10\%.  The MW fraction is $\approx$40\% larger, but consistent at the 1.2$\sigma$ level.   The uniformity of  ${\cal F}_{\rm deep}$ demonstrates that the eclipsing binary fraction of early-B stars does not vary with metallicity beyond the observational uncertainties.  

  Extending toward medium eclipse depths, the values of ${\cal F}_{\rm med}$ in Table 1 are not as undeviating.  Although the MW and LMC OGLE-III samples match within the uncertainty of $\approx$20\%, the OGLE${\mbox -}$II fractions for both the LMC and SMC are statistically lower.  We can resolve this discrepancy by investigating the observed primary eclipse depth distributions ${\cal O}_{\Delta {\rm m}}$($\Delta$m)\,d($\Delta$m), which we display in Figure~1.  The distributions are normalized to the total number of early-B stars so that ${\cal F}_{\rm deep}$~=~$\int_{0.25}^{0.65}$~${\cal O}_{\Delta {\rm m}}$($\Delta$m)\,d($\Delta$m), and the plotted errors {\Large $\sigma$}$_{{\cal O}_{\Delta {\rm m}}}$($\Delta$m) derive from Poisson statistics.  The OGLE-II LMC and SMC data become incomplete at $\Delta$m~$<$~0.25 due to the lower photometric precision of the survey, which leads to the underestimation of ${\cal F}_{\rm med}$.  However, ${\cal O}_{\Delta {\rm m}}$ for all four samples are consistent with each other across the interval for deep eclipses 0.25 $<$ $\Delta$m $<$ 0.65, demonstrating again that the close binary properties of early-B stars do not strongly depend on metallicity.  Using the large and complete LMC OGLE-III sample for eclipse depths  0.10~$<$~$\Delta$m~$<$~0.65, we fit a simple power-law to the eclipse depth distribution.  We find ${\cal S}_{\Delta {\rm m}}$\,d($\Delta$m)~$\propto$~($\Delta$m)$^{-1.65\pm0.07}$\,d($\Delta$m), which we display as the dashed black line in Figure~1.  If this distribution extends toward shallower eclipses, then many additional eclipsing systems may be hiding with $\Delta$m~$<$~0.1.  We return to our discussion of incompleteness corrections in the next section when we conduct Monte Carlo simulations.

\begin{figure}[H]
\center
\includegraphics[width=4.0in]{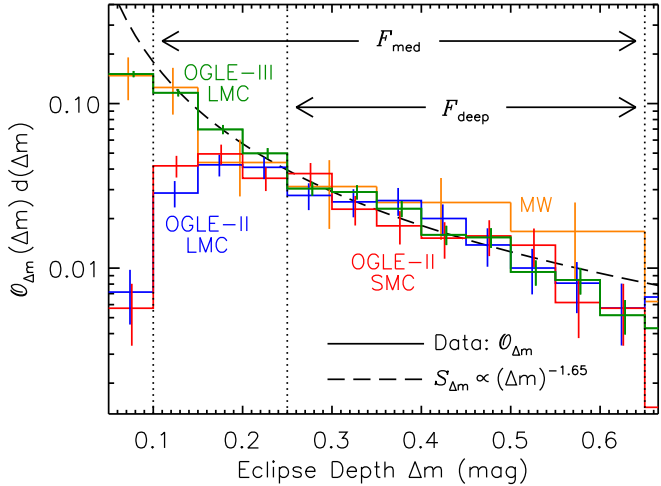}
\caption{The observed primary eclipse depth distribution ${\cal O}_{\Delta {\rm m}}$ with orbital periods $P$~=~2\,$\mbox{-}$\,20~days for early-B stars in the Hipparcos MW (orange), OGLE-II LMC (blue), OGLE-III LMC (green), and OGLE-II SMC (red) samples.   The observed slopes and overall normalizations to ${\cal F}_{\rm deep}$~=~$\int_{0.25}^{0.65}$\,${\cal O}_{\Delta {\rm m}}$($\Delta$m)\,d($\Delta$m)~=~(0.7\,$\mbox{-}$\,1.0)\% of all four samples are consistent with each other across the interval for deep eclipses 0.25 $<$ $\Delta$m $<$ 0.65, demonstrating that the eclipsing binary properties do not substantially change with metallicity.  The OGLE-II data for both the LMC and SMC become incomplete toward shallower eclipses $\Delta$m $\lesssim$ 0.25, while the OGLE-III LMC distribution is relatively complete down to $\Delta$m =  0.10 and is well approximated by a simple power-law ${\cal S}_{\Delta {\rm m}}$ (dashed black).  
}
\end{figure}

In Figure 2, we plot the observed period distributions of eclipsing binaries exhibiting deep eclipses ${\cal O}_{\rm deep}(P)$\,d(log\,$P$) for the three OGLE samples (top panel). We also display the observed period distributions of systems with medium through deep eclipses ${\cal O}_{\rm med}(P)$\,d(log\,$P$) for the complete MW and LMC OGLE-III populations (bottom panel).  Again, we normalize the observed period distributions to the total number of early-B stars so that ${\cal F}_{\rm deep}$ = $\int_{\rm log~2}^{\rm log~20}$\,${\cal O}_{\rm deep}$($P$)\,d(log\,$P$) and ${\cal F}_{\rm med}$~=~$\int_{\rm log~2}^{\rm log~20}$\,${\cal O}_{\rm med}$($P$)\,d(log\,$P$).   The number of eclipsing binaries dramatically increases toward shorter periods, primarily because of geometrical selection effects.  If we ignore limb darkening and tidal distortions, then the probability of eclipses would scale as ${\cal P}$ $\propto$ $P^{-2/3}$ based on Kepler's third law.  If the binaries were distributed uniformly with respect to log $P$ according to \"{O}pik's law \citep{Opik1924, Abt1983}, we would then expect ${\cal S}_{\rm deep}(P)$\,d(log\,$P$)~$\propto$~${\cal S}_{\rm med}(P)$\,d(log\,$P$)~$\propto$~$P^{-2/3}$\,d(log\,$P$).  We display these theoretical curves as the dashed black lines in Figure 2, where the normalization is chosen to guide the eye.  The distributions are shifted slightly toward shorter periods relative to \"{O}pik's prediction, especially the OGLE-II SMC data.  

Although the ${\cal O}_{\rm deep}(P)$ distributions for the OGLE-II and OGLE-III LMC data are consistent with each other, the OGLE-II SMC distribution is discrepantly skewed toward shorter periods.  A K-S test between the OGLE-II LMC and SMC unbinned ${\cal O}_{\rm deep}(P)$ distributions reveals a probability that they derive from the same parent population of only $p_{\rm KS}$ = 0.004.  Similarly, the probability of consistency between the OGLE-II SMC and OGLE-III LMC unbinned ${\cal O}_{\rm deep}(P)$ data is $p_{\rm KS}$~=~0.01.  However, the SMC eclipsing binaries are systematically 0.5 magnitudes fainter, so it is conceivable that some long period systems with shallower eclipses and eclipse durations $\approx$5\% of the total orbital period may have remained undetected in this survey \citep[see][]{Soderhjelm2005}. In fact, we find that all three OGLE samples are consistent with each other, i.e. $p_{\rm KS}$ $>$ 0.1, if we only consider the parameter space of eclipsing binaries with $P$ = 2\,-\,10 days and $\Delta$m = 0.30\,-\,0.65.  We investigate this feature with more robust light curve modeling and Monte Carlo calculations in the next section.

\begin{figure}[H]
\center
\includegraphics[width=5.5in]{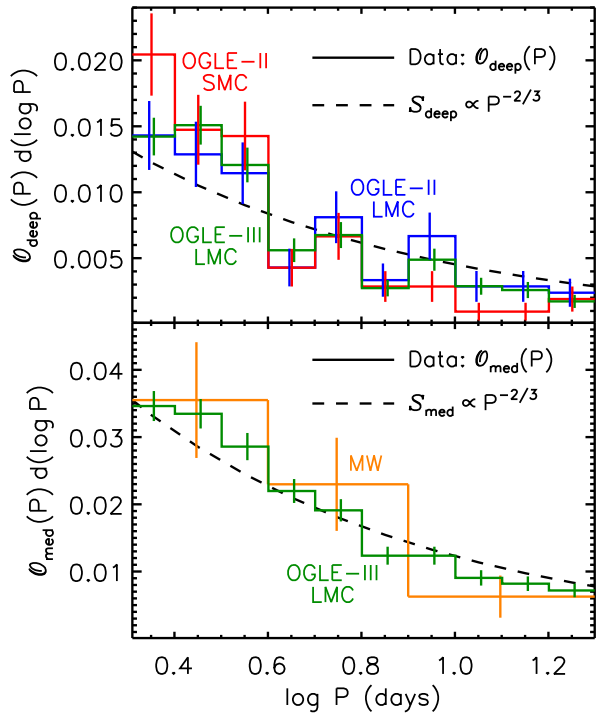}
\caption{The observed orbital period distribution of systems exhibiting deep eclipses ${\cal O}_{\rm deep}$($P$)  (top panel) for the OGLE-II LMC (blue), OGLE-III LMC (green), and OGLE-II SMC (red) samples, and larger population of medium through deep eclipses  ${\cal O}_{\rm med}$($P$) (bottom panel) for the complete MW (orange) and OGLE-III LMC (green) samples.  The distributions are normalized to the total number of early-B stars so that ${\cal F}_{\rm deep}$ = $\int_{\rm log~2}^{\rm log~20}$\,${\cal O}_{\rm deep}$($P$)\,d(log\,$P$)~$\approx$~0.7\% and ${\cal F}_{\rm med}$~=~$\int_{\rm log~2}^{\rm log~20}$\,${\cal O}_{\rm med}$($P$)\,d(log\,$P$)~$\approx$~1.9\%. By making simple approximations and assuming close binaries follow \"{O}pik's law, we would expect ${\cal S}_{\rm deep}(P)$\,d(log\,$P$) $\propto$ ${\cal S}_{\rm med}(P)$\,d(log\,$P$) $\propto$ $P^{-2/3}$\,d(log\,$P$) for the eclipsing binary period distribution (dashed black in both panels).  The observed distributions are weighted toward shorter periods compared to \"{O}pik's prediction, especially the OGLE-II SMC sample.  
}
\end{figure}

\section{Correction for Selection Effects}

We have determined that ${\cal F}_{\rm  deep}$ $\approx$ 0.7\% for all three OGLE samples of eclipsing binaries in the Magellanic Clouds.  The Hipparcos MW value is $\approx$40\% higher, but consistent at the 1.2$\sigma$ level.  Also, both the MW and OGLE-III LMC samples have an observed eclipsing binary fraction with medium eclipse depths of ${\cal F}_{\rm  med}$ $\approx$ 1.9\%.  

In order to make a more stringent comparison, we need to convert the observed eclipsing binary fractions into actual close binary fractions ${\cal F}_{\rm close}$. We define ${\cal F}_{\rm close}$ to be the fraction of systems which have a companion with orbital period 2 days $\le$ $P$ $\le$ 20 days and mass ratio 0.1~$\le$~$q$~$\equiv$~$M_{\rm comp}$/$M_1$~$\le$~1. We must therefore correct for geometrical selection effects and incompleteness toward low-mass companions.   Our ultimate goal is to utilize the observed properties ${\cal O}$ of the eclipsing binary systems, e.g. ${\cal F}_{\rm deep}$ or ${\cal F}_{\rm med}$,  ${\cal O}_{\rm deep}(P)$ or ${\cal O}_{\rm med}(P)$, and ${\cal O}_{\Delta \rm{m}}$($\Delta$m), to derive the underlying properties ${\cal U}$ of the close binary population, e.g. ${\cal F}_{\rm close}$, intrinsic period distribution ${\cal U}_{\rm P}$($P$), and mass-ratio distribution ${\cal U}_{\rm q}(q)$. Although the observational biases of eclipsing binaries have been investigated in the literature \citep[e.g.][]{Farinella1978,Halbwachs1981,Soderhjelm2005}, we wish to conduct detailed modeling specifically suited to our samples in order to accurately quantify the errors.  

   For a given binary with primary mass $M_1$, mass ratio $q$, age $\tau$, metallicity $Z$, and orbital period~$P$, there is a certain probability ${\cal P}$ that the system has an orientation which produces eclipses.  There are even smaller probabilities ${\cal P}_{\rm med}$ and ${\cal P}_{\rm deep}$ that the system has an eclipse depth $\Delta$m which is large enough to be observed in the Hipparcos and OGLE data. We determine these probabilities by first implementing detailed light curve models to compute the eclipse depths $\Delta$m of various binary systems as a function of inclination $i$ (\S3.1).   Using a Monte Carlo technique (\S3.2), we simulate a large population of binaries and synthesize models of the eclipse depth distribution ${\cal M}_{\Delta {\rm m}}$($\Delta$m) and the eclipsing binary period distributions ${\cal M}_{\rm deep}(P)$ and ${\cal M}_{\rm med}(P)$.  We perform thousands of Monte Carlo simulations by making different assumptions regarding the intrinsic period distribution ${\cal U}_{\rm P}$ and mass-ratio distribution ${\cal U}_{\rm q}$.  By minimizing the $\chi^2$ statistic between our Monte Carlo models ${\cal M}$ and observed eclipsing binary data ${\cal O}$, we can determine the probabilities of observing eclipses ${\cal P}_{\rm deep}$ and ${\cal P}_{\rm med}$ as well as the underlying binary properties ${\cal U}$ for each of our populations (\S3.3).  We then account for Malmquist bias in our magnitude-limited samples (\S3.4), and  present our finalized results for ${\cal F}_{\rm close}$ and corrected intrinsic period distribution ${\cal U}_{\rm P}$ (\S3.5). 

\subsection{Light Curve Modeling} 

To simulate eclipse depths $\Delta$m, we use the eclipsing binary light curve modeling software \textsc{nightfall}\footnote{http://www.hs.uni-hamburg.de/DE/Ins/Per/Wichmann/Nightfall.html}.  We incorporate many features of this package, including a square-root limb darkening law, tidal distortions, gravity darkening, model stellar atmospheres, and three iterations of mutual irradiation between the two stars.  For the majority of close binaries with $P$~=~2\,-\,20~days, tides have partially or completely synchronized the orbits as well as dramatically reduced the eccentricities \citep{Zahn1977}, so we assume synchronous rotation and circular orbits in our models.  Nonetheless, several early-B primaries with companions at $P$~=~2\,-\,20~days have measurable non-zero eccentricities, some as large as $e$ $\approx$ 0.6  \citep{Pourbaix2004}. We therefore estimate the systematic error in our determination of ${\cal F}_{\rm close}$ due to the few binaries with these moderate eccentricities (\S3.1.3).   Magnetic bright spots on the surface of massive stars are expected to produce small 10$^{-3}$ mag variations over short durations of days \citep{Cantiello2011}.  Because OGLE and Hipparcos observed the eclipsing binaries over a much longer timespan of years with less photometric precision, we can ignore the effects of starspots.  We compute the \textsc{nightfall} models without any third light contamination, but consider the effects of triple star systems and stellar blending in the crowded Magellanic Cloud OGLE fields using a statistical method (\S3.1.4).    We now synthesize eclipse depths $\Delta$m for the OGLE Magellanic Clouds (\S3.1.1) and Hipparcos MW (\S3.1.2) samples.

\subsubsection{Magellanic Clouds}

To model the OGLE eclipsing binaries, we utilize the Z=0.004, Y=0.26 stellar tracks from the Padova group \citep{Bertelli2008,Bertelli2009}, which correspond to a metallicity between the SMC and LMC mean values.   In addition to basic parameters such as radii $R$($\tau$) and photospheric temperatures $T$($\tau$) as a function of stellar age $\tau$, we also extract the surface gravities $g$($\tau$) from the stellar tracks in order to select appropriate model atmospheres in \textsc{nightfall}. We convert stellar radii to Roche lobe filling factors according to the volume-averaged formula given by \citet{Eggleton1983}.  Although \textsc{nightfall} defines the Roche lobe filling factor along the polar axis, it is more appropriate to use the \citet{Eggleton1983} approximation in cases where the star fills a large fraction of its Roche lobe and is therefore distorted along this potential.  In any case, the volume-averaged Roche lobe radius is only $\approx$\,7\% larger than the polar Roche lobe radius for systems in our sample, so any systematics due to using the \citet{Eggleton1983} formula as input are small.  Based on the numerical calculations performed by \citet{Claret2001} and his comparison to empirical results, we choose an albedo of A~$=$~1.0 for our primary and secondaries hotter than $T$~$>$~7,500K with radiative envelopes ($M_2$~$\ge$~1.3\,\Msun), and A~$=$~0.75 for low-mass secondaries  ($M_2$~$<$~1.3\,\Msun)  at lower temperatures with convective atmospheres.

 Because we selected the OGLE samples from a narrow range of absolute magnitudes,  we can assume that all eclipsing binaries have the same primary mass.  If the luminosity of the primary is dominant, then the median absolute magnitude of $M_{\rm I}$ $\approx$ $-$2.1 in the OGLE samples corresponds to a primary mass of $M_1$ = 12\,\Msun, where we have interpolated the stellar tracks from \citet{Bertelli2009} at half the MS lifetimes as well as utilized bolometric corrections and color indices from \citet{Cox2000}.  However, if the typical secondary in the observed eclipsing systems increases the brightness by $\Delta$M$_{\rm I}$~$\approx$~0.3 mag (see \S3.4.2), then the primary's absolute magnitude of M$_{\rm I}$ $\approx$ $-$1.8 corresponds to $M_1$~=~10\,\Msun.  We therefore adopt $M_1$~=~10\,\Msun\ for all primaries in our simulations.  

We must still consider the systematic error in ${\cal F}_{\rm close}$ due to this single-mass primary approximation.  The sample distributions of absolute magnitudes M$_{\rm I}$ have a dispersion of $\sigma_{\rm M_{\rm I}}$ $\approx$ 0.4 mag, which implies a dispersion in $M_1$ of $\approx$25\%.    According to the mass-radius relation $R$~$\propto$~$M^{0.6}$ and Kepler's law $a$~$\propto$~$M^{1/3}$, then the probability of observing eclipses ${\cal P}_{\rm deep}$ $\propto$ ${\cal P}_{\rm med}$ $\appropto$ $R$/$a$ $\appropto$  $M_1^{0.3}$ due to geometrical selection effects is only weakly dependent on $M_1$.   The systematic error in our derived ${\cal F}_{\rm close}$ = ${\cal F}_{\rm deep}/\langle{\cal P}_{\rm deep}\rangle$ = ${\cal F}_{\rm med}/{\langle\cal P}_{\rm med}\rangle$ is therefore only a factor of 7\% due to the observed dispersion in primary absolute magnitudes $\sigma_{\rm M_{\rm I}}$ $\approx$ 0.4 mag. Similarly, the extinction distributions toward young stars in the Magellanic Clouds have a dispersion of $\sigma_{\rm A_{\rm V}}$ $\approx$ 0.3 mag \citep{Zaritsky1999, Zaritsky2002}, and the I-band excess distributions from the eclipsing companions have a dispersion of $\sigma_{\Delta{\rm M}_{\rm I}}$ $\approx$ 0.2 mag (see \S3.4.2).  These effects contribute additional systematic error factors in ${\cal F}_{\rm close}$ of 6\% and 4\%, respectively.  By adding these three sources of uncertainty in quadrature, we find the total systematic error in ${\cal F}_{\rm close}$ is only a factor of 10\% due to our single-mass primary approximation.  In our estimate for ${\cal P}_{\rm deep}$ $\propto$ ${\cal P}_{\rm med}$ $\appropto$  $M_1^{0.3}$, we have assumed the mass-ratio distributions, and therefore the {\it slopes} of the eclipse depth distributions, do not substantially vary across our narrowly selected interval of primary masses. In fact, for the OGLE-III LMC medium eclipse depth sample, we find ${\cal S}_{\Delta {\rm m}}$~$\propto$~($\Delta$m)$^{-1.54\,\pm\,0.12}$ for the 563 eclipsing binaries brighter than $M_{\rm I}$~=~$-$2.3, and  ${\cal S}_{\Delta {\rm m}}$~$\propto$~($\Delta$m)$^{-1.74\,\pm\,0.11}$  for the 738 systems fainter than $M_{\rm I}$ = $-$2.3.  The consistency of these two slopes justifies our approximation, and therefore our assessment of the systematic error in ${\cal F}_{\rm close}$ is valid.

Because we restricted our samples to observed colors V$-$I $<$ 0.1, i.e. $T_1$ $\gtrsim$ 10,000\,K once reddening is taken into account, most primaries are relatively unevolved on the MS.  For example, a Z = 0.004, $M_1$ = 10\,\Msun\ primary evolves from $R_1$ = 3.3\,\Rsun, $T_1$ = 28,000\,K on the zero-age MS (luminosity class V) to $R_1$ = 8.5\,\Rsun, $T_1$ = 22,000\,K  at the top of the MS by age $\tau_{\rm MS}$~=~23~Myr (technically luminosity class III).  The star then rapidly expands and cools, passing from $R_1$~=~9.0\,\Rsun\ to T$_1$ = 10,000\,K in $\delta t$ $\approx$ 30,000 yrs.  Considering $\delta t$\,/\,$\tau_{\rm MS}$ $\sim$ 10$^{-3}$, the contamination by the few, short-lived bona fide giants with $\tau$ $>$ $\tau_{\rm MS}$ is negligible. 

We calculate the I-band light curve at 1\% phase intervals across the orbit, where we include the effects of fractional visibility of surface elements computed by \textsc{nightfall}.  Because the OGLE eclipsing binary catalogs reported eclipse depths in the I-band as the difference between the dimmest and mean out-of-eclipse magnitudes, we set the zero point magnitude in the \textsc{nightfall} models to the mean value across the phase interval 0.2\,-\,0.3.  We display some example light curves in Figure~3.  The three panels represent orbital periods of $P$ = 2, 6.3, and 20 days, while the colors distinguish various mass ratios $q$ $=$ $M_2$/$M_1$.  We compute the light curves at inclinations $i$ = 77.3$^{\rm o}$, 84.1$^{\rm o}$, and 87.3$^{\rm o}$ from left to right so that the projected separations $a_{\rm proj}$ $\propto$ $P^{2/3}$cos$\,i$ = constant. For spherical stars, the eclipse depths should therefore be identical across these three panels for the same mass ratios.  We evaluate these example models at  age $\tau$ = 17 Myr when the primary reaches an intermediate radius of $R_1$~=~5.3\,\Rsun.

\begin{figure}[H]
\center
\includegraphics[trim = 0cm  3.0cm 0cm 3.0cm, clip, width=6.5in]{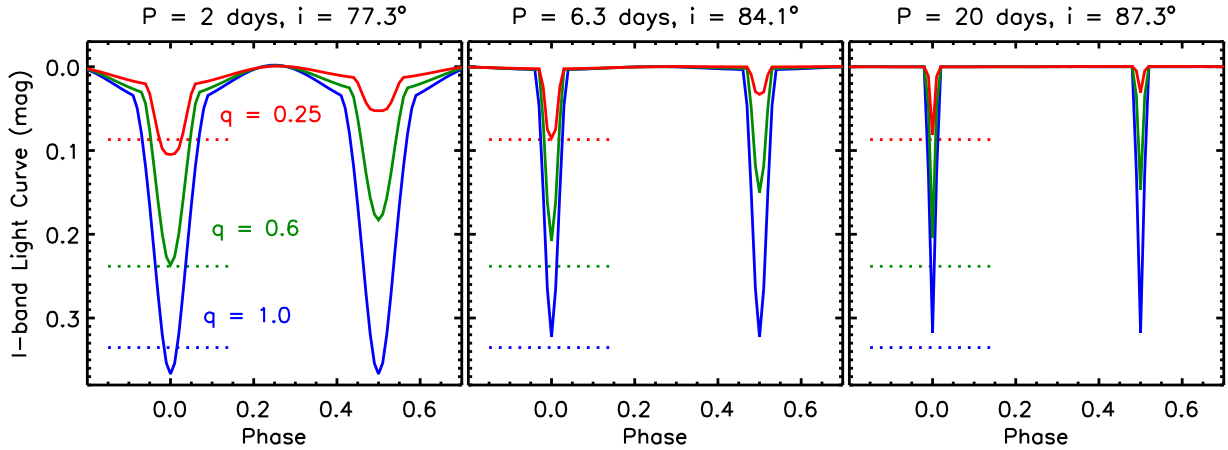}
\caption{Simulated I-band light curves as a function of orbital phase computed by \textsc{nightfall} for various mass ratios $q$ $=$ $M_2$/$M_1$ (distinguished by colors).  The left, middle, and right panels correspond to orbital periods of $P$ = 2, 6.3, and 20 days, respectively, and at the listed inclinations~$i$ which satisfy $P^{2/3}$cos$\,i$ = constant.  All models are evaluated with primary mass $M_1$~=~10\,\Msun\ at age $\tau$ = 17 Myr when $R_1$ = 5.3\,\Rsun.   We compare the detailed  \textsc{nightfall} light curves to simplistic estimates of the maximum eclipse depths which ignore tidal distortions, limb darkening, and color dependence (horizontal dotted lines centered on primary eclipse).  The detailed \textsc{nightfall} models differ from the simple estimates by 0.00\,-\,0.04 mag for these systems, but can reach up to 0.16~mag for older, short-period binaries nearly filling their Roche lobes.  }
\end{figure}

The left panel of Figure 3 with $P$ = 2 days corresponds to primaries filling 60-80\% of their Roche lobes, depending on the mass ratio. The light curves of these close binaries exhibit pronounced ellipsoidal modulations, while the out-of-eclipse magnitudes of systems at longer orbital periods are relatively constant.  In the right panel with $P$ = 20 days, the narrow eclipse widths of 4\% are just at the detectability limit of ecliping binary identification algorithms \citep{Soderhjelm2005}.

  A simple estimate for the eclipse depths can be derived by calculating the bolometric flux in the eclipsed area of the primary assuming spherical stars and no limb darkening.  We compare the \textsc{nightfall} models to this simple approximation for the maximum eclipse depth (horizontal dotted lines centered on primary eclipses).   For $P$ = 2 days, the actual eclipse depths determined by \textsc{nightfall} are generally deeper than the simple approximations because tidal distortions and reflection effects enhance the light curve amplitudes.  Alternatively,  the \textsc{nightfall} results for longer period systems at $P$ = 6.3 and 20 days are typically shallower than the simple approximations because the actual flux eclipsed along grazing angles is less due to the effect of limb darkening.

 Because the OGLE eclipsing binary catalogs exclude ellipsoidal variables that did not exhibit genuine eclipses, we consider only systems with inclinations $i$ $>$ $i_{\rm crit}$ $\equiv$ cos$^{-1}$([$R_1 +R_2$]/a).  We use \textsc{nightfall} to produce a dense grid of eclipse depths $\Delta$m($\tau$, $q$, $P$, $i$) in our parameter space of stellar ages $\tau$ = [0, $\tau_{\rm MS}$ = 23 Myr], mass ratios $q$ = [0.1, 1], orbital periods $P$(days) = [2, 20], and inclinations $i$ = [$i_{\rm crit}$, 90$^{\rm o}$].  In the three panels of Figure 4, we plot our simulated $\Delta$m as a function of inclination $i$ for the same three orbital periods, various mass ratios indicated by color, and for the same $\tau$ = 17 Myr that gives $R_1$ = 5.3\,\Rsun. 

\begin{figure}[H]
\center
\includegraphics[trim = 0cm  2.9cm 0cm 3.0cm, clip, width=6.5in]{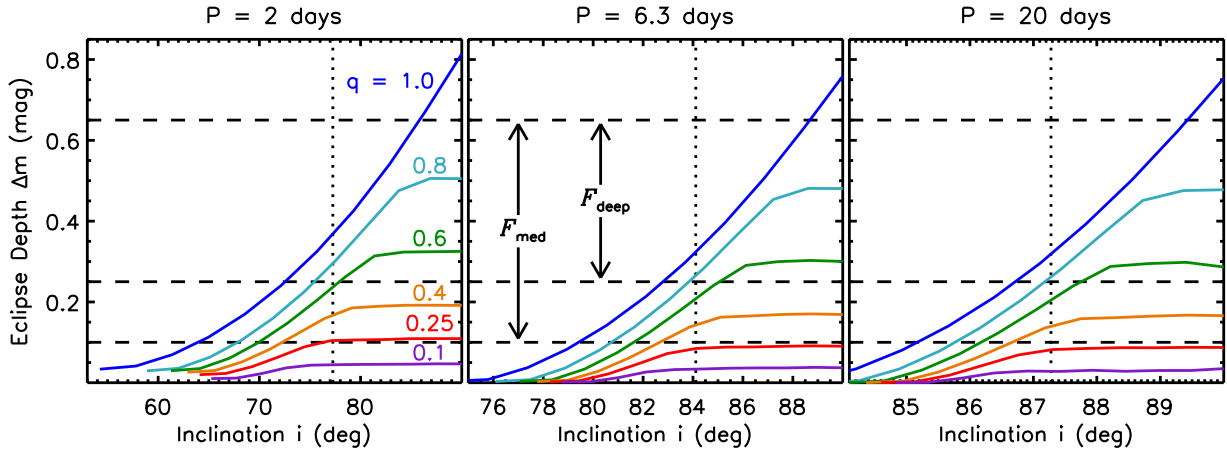}
\caption{Maximum eclipse depths $\Delta$m as a function of inclination $i$ $>$ $i_{\rm crit}$ $\equiv$  cos$^{-1}$([$R_1 +R_2$]/a)  computed using \textsc{nightfall} for various mass ratios $q$ $=$ $M_2$/$M_1$ (distinguished by colors) and three orbital periods (different panels).  We compute the models with the same primary mass $M_1$, age $\tau$, and three orbital periods $P$ as in Figure 3, where the vertical dotted lines represent the inclinations of the systems used to display the light curves.  We also indicate our adopted definition for deep eclipses and the extension toward medium eclipse depths (horizontal dashed lines). The range of inclinations which produce observable eclipses decreases with increasing $P$ simply due to geometrical selection effects.}
\end{figure}

The short-period systems in the left panel of Figure 4 are significantly affected by tidal distortions.  The twin system with $q$ = 1 observed edge-on at $i$ = 90$^{\rm o}$  exceeds the maximum eclipse depth limit for spherical stars of $\Delta$m = 0.75.  Ellipsoidal variables which barely miss eclipses with $i$ =  $i_{\rm crit}$ all have light curve amplitudes of $\Delta$m $<$ 0.05 for this set of parameters (see where curves terminate at bottom left).  For systems which do not fill their Roche lobes, all ellipsoidal variables with $i$ = $i_{\rm crit}$ have amplitudes $\Delta$m $<$ 0.09.  Granted, some systems with $i$ $>$ $i_{\rm crit}$ may not have strong enough eclipse features to be included in the catalog of eclipsing binaries.  Nevertheless, this transition between ellipsoidal variability and genuine eclipses occurs at $\Delta$m $\lesssim$ 0.1, so we can be assured that very few eclipsing systems with measured amplitudes $\Delta$m $>$ 0.1 have been excluded from the catalogs. 

The middle and right panels of Figure 4 represent progressively longer orbital periods where tidal distortions and reflection effects become negligible.   Note the smaller range of inclinations which produce observable eclipses, simply due to geometrical selection effects.  We display with horizontal dashed lines our adopted intervals for deep eclipses and extension toward medium eclipse depths.  Assuming the middle panel is most representative of close binaries with $P$ = 2 - 20 days, then  $i$ $>$ 85$^{\rm o}$ and $q$ $>$ 0.55 are required to observe deep eclipses.  Given random inclinations, the correction factor for geometrical selection effects alone is ${\cal C}_{\rm deep,i}$ $\approx$ $\frac{90^{\rm o}}{90^{\rm o}-85^{\rm o}}$ $\approx$ 18.  Assuming a uniform mass-ratio distribution over the interval $q$~=~[0.1,~1.0], the correction factor for incompleteness toward low-mass companions alone is ${\cal C}_{\rm deep,q}$ $\approx$ $\frac{1-0.1}{1-0.55}$ $\approx$ 2.  The overall probability of observing a system with a deep eclipse is therefore $\langle{\cal P}_{\rm deep}\rangle$ $=$ (${\cal C}_{\rm deep,i}$\,$\times$\,${\cal C}_{\rm deep,q}$)$^{-1}$ $\approx$ 0.03.    Similarly, $i$~$>$~83$^{\rm o}$ and $q$ $>$ 0.3 are required to observe eclipses with medium depths,  implying ${\cal C}_{\rm med,i}$ $\approx$ 13, ${\cal C}_{\rm med,q}$~$\approx$~1.3, and $\langle {\cal P}_{\rm med} \rangle$ $\approx$ 0.06. These two overall probabilities imply similar close binary fractions of ${\cal F}_{\rm close}$~$=$~${\cal F}_{\rm deep}$\,/\,$\langle{\cal P}_{\rm deep}\rangle$ = 0.7\%\,/\,0.03 $\approx$ 25\% and ${\cal F}_{\rm close}$ $=$ ${\cal F}_{\rm med}$\,/\,$\langle{\cal P}_{\rm med}\rangle$ = 1.9\%\,/\,0.06 $\approx$ 30\%. We obtain more precise values in \S3.3 by fitting the observed eclipse depth and period distributions to constrain the actual binary properties.

In Figure 5, we display simulated eclipse depths from \textsc{nightfall} similar to Figure 4, but for constant $P$ = 2.9 days and three different stages of evolution.  The left panel corresponds to zero-age MS systems where the primary radius is $R_1$ = 3.3\,\Rsun, the middle panel represents an intermediate age binary when $R_1$ = 5.3\,\Rsun, and the right panel is for the top of the primary's MS with $R_1$~=~8.5\,\Rsun.  For young systems, $q$ = 0.1 is just at the detectability threshold in our medium eclipse depth samples, which is the primary reason we set the lower limit of our mass-ratio interval to this value.  With increasing $\tau$ and $R_1$,  the range of inclinations which produce visible eclipses increases due to geometrical selection effects.  However, the depths of eclipses for $q$ $\lesssim$ 0.9 become smaller because the fractional area of the primary that is eclipsed decreases with increasing primary radius.  Therefore, our samples of eclipsing binaries are rather incomplete toward smaller, low-mass companions.  For young systems, the probability of observing a low-mass secondary is low, while for older systems the eclipse depths produced by low-mass companions are below the sensitivity of the surveys. 

\begin{figure}[H]
\center
\includegraphics[trim = 0cm  2.9cm 0cm 3.0cm, clip, width=6.5in]{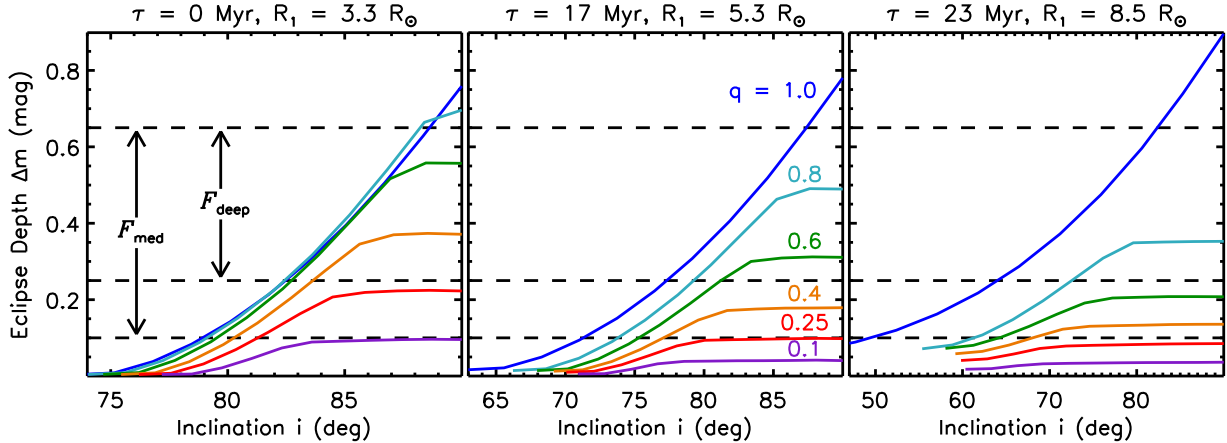}
\caption{Similar to Figure 4, but for the same orbital period of $P$ = 2.9 days and for three different ages $\tau$ on the MS of a 10\,\Msun\ primary. Note the range of inclinations which produce observable eclipses increases with increasing age, while for $q$ $\lesssim$ 0.9 the eclipse depths diminish with age.}
\end{figure}

There is a narrow corner of the parameter space with $P$ $\lesssim$ 2.6 days and $R_1$ $\gtrsim$ 7.0\,\Rsun\ where the primary overfills its Roche lobe.  We assume that either merging or onset of rapid mass transfer causes these systems to evolve outside the parameter space 0.1 $<$ $\Delta$m $<$ 0.65.  In our Monte Carlo simulations (\S3.2), we include their contribution toward the close binary fraction, but remove these systems as eclipsing binaries when fitting ${\cal O}_{\Delta {\rm m}}$($\Delta$m) and either ${\cal O}_{\rm deep}(P)$ or ${\cal O}_{\rm med}(P)$.  A 10\,\Msun\ star spends 8\% of its MS evolution with $R_1$ $>$ 7.0\,\Rsun, and (20\,-\,30)\% of the eclipsing binaries in our samples have orbital periods $P$ $<$ 2.6 days, depending on the survey.   Therefore, the systematc error in our evaluation of the close binary fraction due to these few evolved, close, Roche-lobe filling binaries is only 2\%.  

 For systems which produce eclipse depths $\Delta$m $>$ 0.25 and are not filling their Roche lobes, the root-mean-square deviation between the detailed \textsc{nightfall} simulations and simple approximations which ignore limb darkening and tidal distortions is $\langle \delta$($\Delta$m)$\rangle$ $=$ 0.05~mag.  The difference reaches a maximum value of 0.16~mag for a close period, evolved twin system with $q$ = 1 which nearly fills its Roche lobes. Because of these measurable systematics, it is important that we incorporate the  \textsc{nightfall} results instead of relying on the simple estimates.

\subsubsection{Milky Way}

  We repeat our procedure to model eclipse depths $\Delta$m for the Hipparcos MW sample of eclipsing binaries, but with some slight modifications.  We still assume all primaries have $M_1$ = 10\,\Msun\ because the mean spectral type of our sample is B2, but implement the solar metallicity Z=0.017, Y=0.26 tracks from the Padova group \citep{Bertelli2008, Bertelli2009}.  A solar-metallicity 10\,\Msun\ star has a slightly longer lifetime of $\tau_{\rm MS}$ $\approx$ 25 Myr, and more importantly is (15\,-\,25)\% larger depending on the stage of evolution.  The primary radius is $R_1$\,=\,3.8\,\Rsun\ on the zero-age MS versus $R_1$ = 3.3\,\Rsun\ for the Z=0.004 model, and reaches $R_1$\,=\,10.5\,\Rsun\ at the top of the MS compared to $R_1$ = 8.5\,\Rsun\ for the low-metallicity track.  For the same close binary properties, we actually expect ${\cal F}_{\rm deep}$ in the MW to be 20\% higher because the probability of eclipses scales as ${\cal P}$ $\propto$ ($R_1$ + $R_2$).  This radius-metallicity relation diminishes the already small 1.2$\sigma$ difference between the MW and Magellanic Cloud statistics inferred from ${\cal F}_{\rm deep}$.  Finally, we evaluate the eclipse depth $\Delta$m based on the V-band light curves computed by \textsc{nightfall}, which closely approximates the Hipparcos passband.  

\subsubsection{Eccentric Orbits}

 Because we attempt to address all sources of error, we account for systematics due to eccentric orbits in our determination of the close binary fraction. Unfortunately, the extent of tidal distortions and mutual irradiation continually change in an eccentric orbit, so that all the binary properties in \textsc{nightfall} must be recalculated at each phase of the orbit instead of solely varying the orientation.  It would become computationally too expensive if we were to add dimensions of eccentricity $e$ and periastron angle $\omega$ to our original grid of models $\Delta$m($\tau$, $q$, $P$, $i$).

Since the eccentricities of close binaries are relatively small due to tidal circularization, we can determine the average error $\langle \delta$($\Delta$m)$\rangle$ of a representative eclipsing binary and propagate the uncertainty into our evaluation of ${\cal F}_{\rm close}$.  We consider an eclipsing binary with $\tau$ = 17\,Myr, $q$~=~0.6, $P$ = 4 days, and $i$ = 90$^{\rm o}$ as our test example, which gives $\Delta$m = 0.30 for a circular orbit (see Figure 4). For the 101 systems in the ninth catalog of spectroscopic binary orbits \citep{Pourbaix2004} with measured eccentricities, orbital periods $P$ = 2\,-\,20\,days, and primaries with spectral types B0-B3 and luminosity classes III-V, the average eccentricity is only $\langle e \rangle$ = 0.17. For the 56 systems with $P$ = 2\,-\,5\,days, which is more representative of our eclipsing binary sample, the mean eccentricity is even lower at $\langle e \rangle$ = 0.11. Using \textsc{nightfall}, we calculate the eclipse depths for our test example at an intermediate value of $e$ = 0.15 as well as an upper $\approx$1$\sigma$ value of $e$ = 0.30, each at varying periastron angles $\omega$.  

For $e$ = 0.15, we find the eclipse depths vary by $\delta$($\Delta$m) $<$ 0.004 mag compared to a circular orbit, with an average value of $\langle\delta$($\Delta$m)$\rangle$ = 0.002 mag if we weight uniformly with respect to $\omega$.  The error is slightly higher at $\langle\delta$($\Delta$m)$\rangle$ = 0.005 mag for $e$ = 0.30.  We found in \S3.1.1 that the average error between the detailed \textsc{nightfall} models and simple estimates ignoring tidal distortions and limb darkening is $\langle\delta$($\Delta$m)$\rangle$ = 0.05 mag.  We show in \S3.3 that this would have propagated into a systematic error factor of 20\% in our determination of ${\cal F}_{\rm close}$.  Since the error in eclipse depths due to eccentric orbits is an order of magnitude smaller, we expect the uncertainty in ${\cal F}_{\rm close}$ due to non-circular orbits to be only a factor of 2\%.

\subsubsection{Third Light Contamination}

  A third light source can have a much larger effect on the observed eclipse depth $\Delta$m of an eclipsing binary, depending on the luminosity of the contaminant.  We first consider wider companions in triple star systems. About 40\% of early-type primaries have a visually resolved companion \citep{Turner2008,Mason2009}. More importantly, most close binaries, such as our eclipsing systems, are observed to be the inner components of triple star systems \citep{Tokovinin2006}.  Specifically, this study found that 96\% of binaries with $P$ $<$ 3 days have a wider tertiary companion.  Assuming the typical eclipsing secondary increases the brightness by $\Delta$M = 0.3 mag (see \S3.4), then a tertiary companion with $q$~$=$~$M_3$/$M_1$~$>$~0.5 is capable of increasing the system luminosity by $\gtrsim$10\%.  The wider companions around early-type primaries are observed to be drawn from a mass-ratio distribution weighted toward lower mass, fainter stars \citep{Abt1990,Preibisch1999, Duchene2001, Shatsky2002}. These observations find that only (10\,-\,30)\% of wide companions have mass ratios $q$ $>$ 0.5.  Even if every eclipsing binary has one wider component, we would expect that only $\approx$20\% of tertiaries have large enough luminosities to measureably affect our light curve modeling.  

  We also consider third light contamination due to stellar blending in the crowded Magellanic Cloud fields.  Based on the OGLE photometric catalogs, there are 4.2 million \citep{Udalski2000}, 12 million \citep{Udalski2008}, and 1.5 million \citep{Udalski1998} systems with M$_{\rm I}$ $>$ 1.2 in the OGLE-II LMC, OGLE-III LMC, and OGLE-II SMC footprints, respectively.    The median absolute magnitude of these sources is $M_{\rm I}$ $\approx$ 0.4, which is 10\% the I-band luminosity of our median early-B eclipsing binary with M$_{\rm I}$~$\approx$~$-$2.1. The average space densities of stars with M$_{\rm I}$ $>$ 1.2 are 0.07, 0.03, and 0.05 objects per square arcsecond in the OGLE-II LMC, OGLE-III LMC, and OGLE-II SMC fields, respectively.  Given a median seeing of 1.2$''$-1.3$''$ during the OGLE observations, we expect only (5\,-\,12)\% of early-B eclipsing binaries to be blended with sources brighter than  M$_{\rm I}$ = 1.2.  The probability of stellar blending with a background/foreground source is slightly smaller than the probability of contamination in a triple star system, where in both cases we included third light components $\gtrsim$10\% the luminosity of the eclipsing system.

  Because a sizable fraction of eclipsing binaries are affected by third light contamination from stellar blending and triples systems, we model the third light sources in the eclipsing binary populations using a statistical method.  When we conduct our Monte Carlo simulations in the next section, we synthesize distributions of eclipse depths $\Delta$m based on our \textsc{nightfall} models, but assume that a 20\% random subset of eclipsing systems have reduced eclipse depths $\Delta$m$_{\rm measured}$ = 0.8\,$\Delta$m$_{\rm true}$.   These values approximate the probabilities and representative luminosities of the third light contaminants.  By comparing our model fits with and without the third light sources, we can gauge the effect on our derived close binary properties.  

\subsection{Monte Carlo Simulations}

   The eclipsing binary samples provide the distributions of observed orbital periods and eclipse depths.  We would like to use this information to learn as much as possible about the properties of the close binary populations in the different environments.  To do this, we use the fact that the eclipse depths $\Delta$m($M_1$, $q$, $Z$, $\tau$, $P$, $i$) are determined by six physical properties of the binary.  Based on our single-mass approximation discussed in \S3.1.1, we only consider $M_1$ = 10\Msun\ primaries and propagate the systematic error from this approximation into our finalized results for the close binary fraction.  We also evaluate our models for two main metallicity groups: one using the Z=0.004 stellar tracks and I-band eclipse depths to be compared to the three OGLE Magellanic Cloud samples, and one using the Z=0.017 stellar tracks and V-band eclipse depths to be compared to the Hipparcos MW data.  The four remaining binary properties $\tau$, $i$, $P$, and $q$ are characterized by the distribution functions below, some of which have one or more free parameters $\vec{x}$. To simulate a population of binaries, we use a random number generator to select systems from these distribution functions.  We then conduct a set of Monte Carlo simulations, where each simulation is characterized by a particular combination of model parameters $\vec{x}$.

 Because the star formation rates of the Magellanic Clouds \citep{Indu2011} and local solar neighborhood in the MW \citep{delaFuenteMarcos2004} have not dramatically changed over the most recent $\tau_{\rm MS}$ $\approx$ 24~Myr, we select 10\,\Msun\ primaries from a uniform age distribution across the interval $\tau$~=~[0,~$\tau_{\rm MS}$]. The close binary fraction ${\cal F}_{\rm close}$ is one of the free parameters $\vec{x}$, and for each binary, we assume random inclinations in the range $i$~=~[0$^{\rm o}$,~90$^{\rm o}$].  We select an orbital period from the distribution:

\begin{equation}
 {\cal U}_{\rm P}(P)\,{\rm d}({\rm log}\,P) = {\cal K}_{\rm P}\,P^{\gamma_{\rm P}}\,{\rm d}({\rm log}\,P)
\end{equation}

\noindent across the interval log\,2~$\le$~log\,$P$\,(days)~$\le$~log\,20.  For a given Monte Carlo simulation, we fix the period exponent $\gamma_{\rm P}$, but consider 21 different values in the range $-$1.5~$\le$~$\gamma_{\rm P}$~$\le$~0.5 evaluated at $\Delta$$\gamma_{\rm P}$~=~0.1 intervals when synthesizing different populations of binaries.  Note that \"{O}pik's law gives $\gamma_P$~=~0.  The normalization constant ${\cal K}_{\rm P}$ satisfies ${\cal F}_{\rm close}$~=~$\int_{\rm log~2}^{\rm log~20}{\cal U}_{\rm P}(P)\,$d(log $P$). 

Although the mass-ratio distribution is typically described as a power-law, there is evidence that close binaries harbor an excess fraction of twins with mass ratios approaching unity \citep{Tokovinin2000,Halbwachs2003,Lucy2006,Pinsonneault2006}.  We therefore implement a two-parameter formalism:

\begin{equation}
 {\cal U}_{\rm q}(q)\,{\rm d}q = {\cal K}_{\rm q} \Big[\frac{1-{\cal F}_{\rm twin}}{15}\,e^{\gamma_q}\,q^{\gamma_q} + {\cal F}_{\rm twin}\,q^{15} \Big] {\rm d}q
\end{equation}

\noindent over the interval 0.1~$\le$~$q$~$\le$~1. We consider 36 values for the mass-ratio exponent in the range $-$2.5~$\le$~$\gamma_{\rm q}$~$\le$~1.0 evaluated at $\Delta\gamma_{\rm q}$~=~0.1 intervals, and 16 values for the excess twin fraction in the range 0~$\le$~${\cal F}_{\rm twin}$~$\le$~0.3 at $\Delta{\cal F}_{\rm twin}$~=~0.02 intervals.  Again, the normalization constant ${\cal K}_{\rm q}$ satisfies ${\cal F}_{\rm close}$~=~$\int_{0.1}^{1}{\cal U}_{\rm q}(q)\,$d$q$.  The coefficients in the above equation approximate the relative contribution of the two terms so that the integrated fraction of close binaries in the peak toward unity is ${\cal F}_{\rm twin}$ while the total fraction of close binaries in the low-$q$ tail is 1~$-$~${\cal F}_{\rm twin}$.  

Once we have selected a binary with age $\tau$, inclination $i$, period $P$, and mass ratio $q$, we determine its eclipse depth by interpolating our grid of models $\Delta$m($\tau$, $i$, $P$ $q$).  We simulate 10$^6$ binaries for each combination of parameters $\gamma_{\rm P}$, $\gamma_{\rm q}$, and ${\cal F}_{\rm twin}$, resulting in 21\,$\times$\,36\,$\times$\,16 = 12,096 sets of Monte Carlo simulations.    The fourth free parameter ${\cal F}_{\rm close}$ determines the overall normalization, and we consider 71 different values in the range 0.05~$\le$~${\cal F}_{\rm close}$~$\le$~0.4 evaluated at $\Delta{\cal F}_{\rm close}$~=~0.005 intervals.

For each combination of parameters $\vec{x}$ = \{$\gamma_{\rm P}$, $\gamma_{\rm q}$, ${\cal F}_{\rm twin}$, ${\cal F}_{\rm close}$\}, we synthesize our model distributions ${\cal M}_{\Delta {\rm m}}$($\Delta$m,$\,\vec{x}$), ${\cal M}_{\rm deep}(P,\,\vec{x})$, and ${\cal M}_{\rm med}(P,\,\vec{x})$. For our primary results, we have incorporated the detailed \textsc{nightfall} models where a 20\% random subset have eclipse depths reduced by 20\% in order to account for third light contamination (\S3.1.4).  For comparison, we also evaluate the eclipse depths using the \textsc{nightfall} models without third light contamination as well as using the simple bolometric estimates which ignore tidal distortions and limb darkening.

\subsection{Fitting the Data}

\subsubsection{Mass-ratio Distribution ${\cal U}_{\rm q}$}

We initially fit the observed eclipse depth distribution ${\cal O}_{\Delta {\rm m}}$ only, which primarily constrains the  mass-ratio distribution ${\cal U}_{\rm q}$ as well as the normalization to ${\cal F}_{\rm close}$ according to Eq. 2. We determine the best-fit model parameters $\vec{x}$ = \{$\gamma_{\rm P}$, $\gamma_{\rm q}$, ${\cal F}_{\rm twin}$, ${\cal F}_{\rm close}$\} by minimizing the $\chi^2_{\Delta {\rm m}}$($\vec{x}$) statistic between the observed eclipse depth distribution ${\cal O}_{\Delta {\rm m}}$($\Delta$m) and our Monte Carlo models ${\cal M}_{\Delta {\rm m}}$($\Delta$m, $\vec{x}$):

\begin{equation}
 \chi^2_{\Delta {\rm m}}(\vec{x}) = \sum_k^{N_{\Delta {\rm m}}} \Big(\frac{{\cal O}_{\Delta {\rm m}}(\Delta{\rm m}_k) - {\cal M}_{\Delta {\rm m}}(\Delta{\rm m}_k,\vec{x})}{\mbox{\Large $\sigma$}_{{\cal O}_{\Delta {\rm m}}}(\Delta{\rm m}_k)}\Big)^2 
\end{equation}

\noindent We sum over the bins of data displayed in Figure 6 that are complete, specifically the $N_{\Delta {\rm m}}$~=~8 bins across 0.25 $<$ $\Delta$m\,(mag) $<$ 0.65 for the OGLE-II LMC and SMC populations, $N_{\Delta {\rm m}}$ = 5 bins across 0.10 $<$ $\Delta$m $<$ 0.65 for the MW,  and the $N_{\Delta {\rm m}}$ = 11 bins across 0.10 $<$ $\Delta$m $<$ 0.65 for the OGLE-III LMC sample.  In Figure 6, we display the best-fit models ${\cal M}_{\Delta {\rm m}}$($\Delta$m) for each sample, together with the data. Although we have excluded eclipsing binaries with $\Delta$m $>$ 0.65 mag, which derive from nearly edge-on twin systems as well at evolved binaries that have filled their Roche lobes, twins are most likely to have grazing trajectories that produce eclipse depths in our selected parameter space (see \S3.1.1).  For the OGLE Magellanic Cloud samples that have large sample statistics in the interval 0.40 mag $<$ $\Delta$m $<$ 0.65 mag, we therefore have sufficient leverage to constrain the excess twin fraction.

\begin{figure}[H]
\center
\includegraphics[width=7.2in]{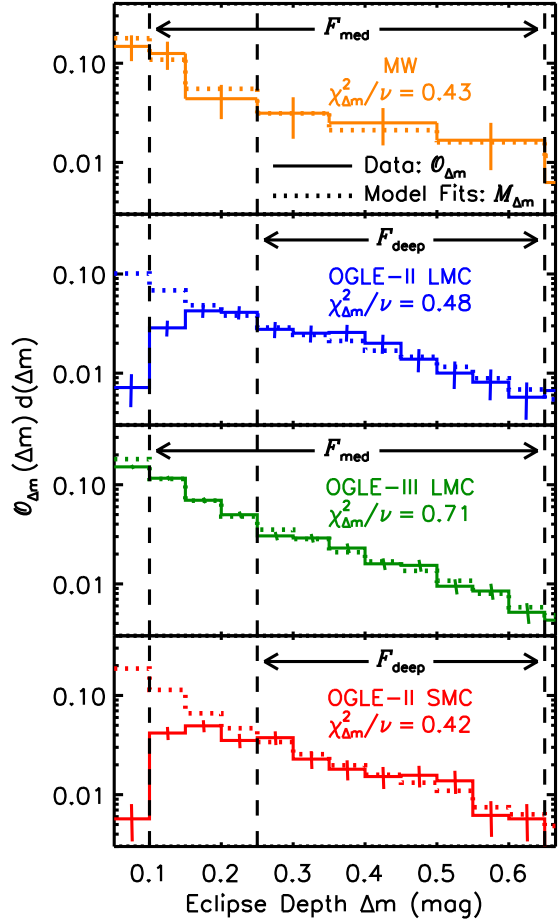}
\caption{The observed primary eclipse depth distributions ${\cal O}_{\Delta{\rm m}}$ (solid) as displayed in Figure 1 for Hipparcos MW (orange), OGLE-II LMC (blue), OGLE-III LMC (green), and OGLE-II SMC (red) populations.  We determine the best-fit Monte Carlo models ${\cal M}_{\Delta {\rm m}}$ (dotted) by minimizing the $\chi^2_{\Delta {\rm m}}$ statistic across the ${\cal F}_{\rm deep}$ interval for the OGLE-II LMC and SMC data and over the ${\cal F}_{\rm med}$ interval for the MW and OGLE-III LMC populations, but we display the full histograms for reference.}
\end{figure}

 The observed eclipse depth distributions can only constrain ${\cal F}_{\rm close}$, $\gamma_{\rm q}$, and ${\cal F}_{\rm twin}$, which effectively gives $\nu$~=~$N_{\Delta {\rm m}}$\,$-$\,3 degrees of freedom.  We report in Table~2 the minimized reduced $\chi^2_{\Delta {\rm m}}$ statistics, degrees of freedom $\nu$, and probabilities to exceed $\chi^2_{\Delta {\rm m}}$.   We calculate a grid of joint probabilities $p_{\vec{x}}(\vec{x})$~$\propto$~$e^{-\chi^2_{\Delta {\rm m}}(\vec{x})/2}$, and then marginalize over the various parameters to calculate the probability density functions $p_{x_i}(x_i)$ for each parameter $x_i$. In Table~2, we list the average values $\mu_{x_i}$\,=\,$\int x_i\,p_{x_i}(x_i)$\,d$x_i$ and uncertainties $\sigma_{x_i}$\,=\,$[\int (x_i - \mu_{x_i})^2\,p_{x_i}(x_i)\,$d$x_i]^{1/2}$ of the three parameters constrained by ${\cal O}_{\Delta {\rm m}}$ for each of the eclipsing binary samples.   Some of the parameters are correlated and have asymmetric probability density distributions, so we display two dimensional probability contours $p_{x_i, x_j}$$(x_i, x_j)$ for some combinations of parameters in Figure 7.

\begin{table}[H]
\small
\center
\begin{tabular}{|c|c|c|c|c|c|c|}
\hline
Sample       & $\chi^2_{\Delta {\rm m}}/{\nu}$ & $\nu$ & PTE  & ${\cal F}_{\rm twin}$ & $\gamma_{\rm q}$   & ${\cal F}_{\rm close}$ \\
\hline
MW           & 0.43                            &   2   & 0.65 & 0.16\,$\pm$\,0.10     & $-$0.9\,$\pm$\,0.8 & 0.22\,$\pm$\,0.06~      \\
\hline
OGLE-II LMC  & 0.48                            &   5   & 0.79 & 0.10\,$\pm$\,0.07     & $-$0.6\,$\pm$\,0.7 & 0.21\,$\pm$\,0.08~      \\
\hline
OGLE-III LMC & 0.71                            &   8   & 0.68 & 0.04\,$\pm$\,0.03     & $-$1.0\,$\pm$\,0.2 & 0.27\,$\pm$\,0.05~      \\ 
\hline
OGLE-II SMC  & 0.42                            &   5   & 0.83 & 0.08\,$\pm$\,0.06     & $-$0.9\,$\pm$\,0.7 & 0.24\,$\pm$\,0.08~      \\
\hline
\end{tabular}
\caption{Results of our Monte Carlo simulations and fits to the observed eclipse depth distributions ${\cal O}_{\Delta {\rm m}}$ {\it only}.  For each of the eclipsing binary samples, we list the minimized reduced $\chi^2_{\Delta {\rm m}}$ statistics, degrees of freedom $\nu$ = $N_{\Delta {\rm m}}$ $-$ 3, probabilities to exceed $\chi^2_{\Delta {\rm m}}$ given $\nu$, and the mean values and 1$\sigma$ uncertainties of the three model parameters constrained by ${\cal O}_{\Delta {\rm m}}$.  }
\end{table}

\begin{figure}[H]
\center
\includegraphics[width=6.5in,trim= 0cm 2.1cm 0cm 1.2cm,clip=true]{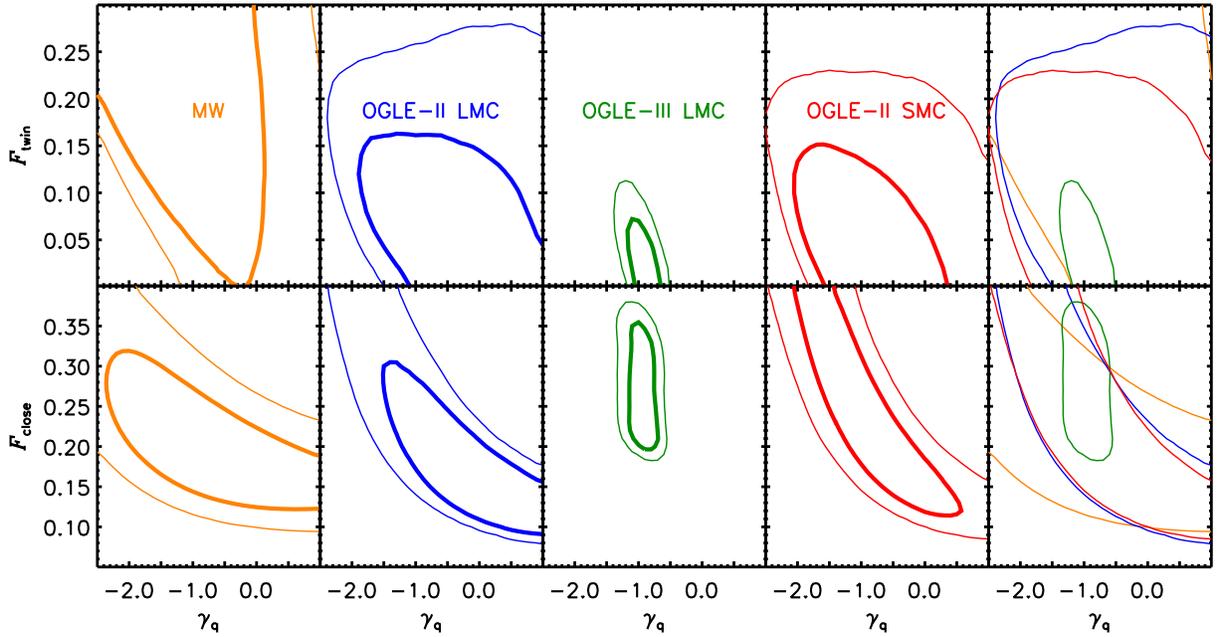}
\caption{Probability contours at the 1$\sigma$ (thick) and 2$\sigma$ (thin) confidence levels of model parameter combinations constrained {\it only} by the observed eclipse depth distributions ${\cal O}_{\Delta {\rm m}}$  for the Hipparcos MW (orange), OGLE-II LMC (blue), OGLE-III LMC (green), and OGLE-II SMC (red) populations. In the top panels,  the OGLE-III LMC data clearly demonstrates a distribution weighted toward lower mass secondaries compared to a uniform distribution with $\gamma_{\rm q}$ = 0, and the other populations also favor negative values for the mass-ratio distribution exponent $\gamma_{\rm q}$.  For the three OGLE Magellanic Cloud samples, we find a small excess twin population with $q$~$\gtrsim$~0.9 of only ${\cal F}_{\rm twin}$~$\approx$~(4\,-\,10)\%. In the bottom panels, all four samples are consistent with a close binary fraction of ${\cal F}_{\rm close}$ $\approx$ 25\% and a mass-ratio distribution exponent of $\gamma_{\rm q}$ $\approx$ $-$1.0.}
\end{figure}

The higher quality OGLE-III LMC population, with its larger sample size and completeness down to $\Delta$m = 0.10, best constrains the model parameters. We find a negligible excess fraction of twins ${\cal F}_{\rm twin}$ = (4\,$\pm$\,3)\%, a mass-ratio distribution weighted toward low-mass companions with $\gamma_{\rm q}$~=~$-$1.0\,$\pm$\,0.2, and a close binary fraction of ${\cal F}_{\rm close}$ = (27\,$\pm$\,5)\% (before corrections for Malmquist bias - see \S3.4).  Based on our Monte Carlo simulations, a uniform mass-ratio distribution would have produced ${\cal S}_{\Delta {\rm m}}$\,d($\Delta$m) $\appropto$ ($\Delta$m)$^{-1.0}$\,d($\Delta$m), not as steep as the observed trend ${\cal S}_{\Delta {\rm m}}$\,d($\Delta$m) $\propto$ ($\Delta$m)$^{-1.65\pm0.07}$\,d($\Delta$m).

The less complete and/or smaller MW, OGLE-II LMC, and OGLE-II SMC samples do not permit precise determinations of $\gamma_{\rm q}$.   Nonetheless, the fitted mean values for these three samples span the range $\gamma_{\rm q}$ = $-0.9$\,-\,$-0.6$, suggesting these binary populations also favor low-mass companions.    For these populations, our solutions for the model parameters ${\cal F}_{\rm close}$ and $\gamma_{\rm q}$ are anti-correlated (see bottom panels of Figure 7).  This is because a larger fraction of low-mass secondaries below the threshold of the survey sensitivity implies a higher ${\cal F}_{\rm close}$ given the same ${\cal F}_{\rm deep}$.  All four samples are consistent with a close binary fraction of ${\cal F}_{\rm close}$ $\approx$ 25\%, which matches our initial estimate in \S3.1.   The precise values will decrease slightly once we correct for Malmquist bias (see \S3.4).

Even though $\gamma_{\rm q}$ is not well known for the OGLE-II data, we can still constrain the excess twin fraction to be ${\cal F}_{\rm twin}$ $\approx$ (4\,-\,10)\% for all three OGLE Magellanic Cloud samples (see top panels of Figure 7).   A dominant twin population would have caused the eclipse depth distribution ${\cal O}_{\Delta {\rm m}}$ to flatten or even rise toward the deepest eclipses $\Delta$m $>$ 0.4.  Instead, the observed eclipse depth distributions for the three OGLE Magellanic Clouds samples  continue with the same power-law ${\cal S}_{\Delta {\rm m}}$~$\propto$~($\Delta$m)$^{-1.65}$.  Because there are very few eclipsing binaries with $\Delta$m $>$ 0.4 in the MW data, we cannot adequately measure ${\cal F}_{\rm twin}$ for this population, but see our well-constrained estimate of ${\cal F}_{\rm twin}$ $\approx$ 7\%  based on spectroscopic observations of early-type stars in the MW (\S4).   

  We have reported fitted parameters based on the \textsc{nightfall} models where a 20\% random subset have eclipse depths reduced by 20\% to account for third light contamination (\S3.1.4).  Because shallower eclipses systematically favor lower mass companions, the fitted mass-ratio distributions would have been shifted toward {\it even lower} values, albeit slightly, had we not considered this effect.   Specifically, we find the excess twin fraction would have decreased by $\Delta{\cal F}_{\rm twin}$~=~0.01$\,\mbox{-}\,$0.03 and the mass-ratio distribution exponent would have decreased by $\Delta\gamma_{\rm q}$ = 0.0$\,\mbox{-}\,$0.2, depending on the sample.  The close binary fraction would have changed by a factor of (3\,-\,6)\%, i.e. $\Delta{\cal F}_{\rm close}$~$\approx$~0.01, with no general trend on the direction. Hence, third light contamination only mildly affects the inferred close binary properties.

\subsubsection{Probabilities of Observing Eclipses ${\cal P}_{\rm deep}(P)$ and ${\cal P}_{\rm med}(P)$}
  
 The probabilities ${\cal P}_{\rm deep}(P)$ and ${\cal P}_{\rm med}(P)$ are defined to be the ratios of systems exhibiting deep (0.25 $<$ $\Delta$m $<$ 0.65) and medium (0.10 $<$ $\Delta$m $<$ 0.65) eclipses, respectively,  to the total number of companions with $q$ $>$ 0.1 at the designated period.  These probabilities obviously decrease with increasing orbital period $P$ due to geometrical selection effects.  In addition, ${\cal P}_{\rm deep}(P)$ and ${\cal P}_{\rm med}(P)$ depend  on the metallicity $Z$, which determines the radial evolution of the stellar components, and also on the underlying mass-ratio distribution ${\cal U}_{\rm q}$.   Mass-ratio distributions which favor lower-mass, smaller companions result in lower probabilities of observing eclipses because a larger fraction of the systems have eclipse depths below the sensitivity of the surveys.  Because we have constrained ${\cal U}_{\rm q}$ for each of the four eclipsing binary populations, we have already effectively determined these probabilities from our Monte Carlo simulations. We use these more accurately constrained probabilities when we account for Malmquist bias in \S3.4 as well as to visualize the corrected period distribution in \S3.5.  

  Using our solutions for ${\cal U}_{\rm q}$  for each of the four eclipsing binary samples, we display the resulting ${\cal P}_{\rm deep}(P)$ and ${\cal P}_{\rm med}(P)$ in Figure 8.  We propagate the fitted errors in $\gamma_{\rm q}$ and ${\cal F}_{\rm twin}$ , as well as their mutual correlation as displayed in the top panels of Figure 7, to determine the uncertainties in the probabilities.  For comparison, we calculate  ${\cal P}_{\rm med}(P)$ and ${\cal P}_{\rm deep}(P)$ assuming the low-metallicity $Z$~=~0.004 stellar tracks and a  uniform mass-ratio distribution ${\cal U}_{\rm q}$, i.e. $\gamma_{\rm q}$ = 0 and ${\cal F}_{\rm twin}$ = 0. 

In the top panel of Figure 8, the probabilities ${\cal P}_{\rm deep}$ for the OGLE Magellanic Cloud samples, which all have fitted values of $\gamma_{\rm q}$ that are negative, are systematically lower than the probabilities which assume a uniform mass-ratio distribution.   Based on our back-of-the-envelope estimates in \S3.1.1 where we assumed a uniform mass-ratio distribution, we determined that the correction factor between ${\cal F}_{\rm deep}$ and ${\cal F}_{\rm close}$ due to incompleteness toward low-mass companions {\it alone} was ${\cal C}_{\rm deep,q}$~$\approx$~2.  The fact that the fitted mass-ratio distributions favor more low-mass companions increases this correction factor to ${\cal C}_{\rm deep,q}$ $\approx$ 3.  Nonetheless, the overall probability of observing deep eclipses at intermediate periods of log\,$P$ = 0.8 is ${\cal P}_{\rm deep}$ = 0.02\,-\,0.04, depending on the model, which spans our estimated average in \S3.1 of  $\langle {\cal P}_{\rm deep} \rangle$ = 0.03.  Finally note the intrinsically small probability of observing deep eclipses at long periods, e.g. only ${\cal P}_{\rm deep}$ $\approx$ 1\% of all binaries at $P$~=~20~days are detectable as eclipsing systems with 0.25 $<$ $\Delta$m $<$ 0.65.

In the bottom panel of Figure 8, the variations in ${\cal P}_{\rm med}$ are significantly smaller.  This is because the probability of observing eclipses becomes less dependent on the underlying mass-ratio distribution as the observations become more sensitive to shallower eclipses.  Essentially, the correction factor for incompleteness toward low-mass companions alone is only ${\cal C}_{\rm med,q}$~=~1.5, slightly larger than our original estimate of ${\cal C}_{\rm med,q}$ = 1.3 in \S3.1.1, but still very close to unity.    The MW correction factor ${\cal C}_{\rm med,i}$ for geometrical selection effects is 20\% smaller than the OGLE-III LMC values, and therefore the overall probabilities ${\cal P}_{\rm med}$ are 20\% larger.  This is consistent with our interpretation of the radius-metallicity relation in \S3.1.2.  \citet{Soderhjelm2005} calculated the probabilities of observing solar-metallicity eclipsing binaries with $\Delta$m $>$ 0.1 as a function of spectral type and period.  Because the fraction of systems with $\Delta$m $>$ 0.65 is negligible compared to the fraction with 0.1 $<$ $\Delta$m $<$ 0.65, we can compare the \citet{Soderhjelm2005} results to our  ${\cal P}_{\rm med}$($P$).   We interpolate the probabilities in their Table A.1 for OB stars with  $\langle$M$_{\rm V} \rangle$~=~$-$3.04 and B stars with $\langle$M$_{\rm V} \rangle$~=~$-$0.55 for our sample's median value of M$_{\rm V}$~$\approx$~$-$2.3.  The resulting ${\cal P}_{\rm med}$, which we display in the bottom panel of Figure 8, is consistent with our MW distribution. At log\,$P$~=~0.8, the OGLE-III LMC value of ${\cal P}_{\rm med}$ = 0.06 matches our initial estimate in \S3.1.1 of $\langle {\cal P}_{\rm med} \rangle$ = 0.06.  

\begin{figure}[H]
\center
\includegraphics[width=5.0in]{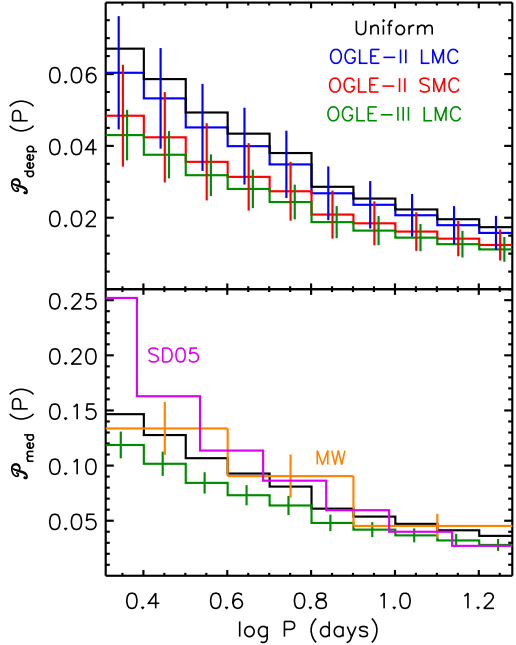}
\caption{The probabilities that a companion with $q$ $>$ 0.1 exhibits deep ${\cal P}_{\rm deep}$ (top) and medium ${\cal P}_{\rm med}$ (bottom) eclipses using our fitted solutions to the overall mass-ratio distribution ${\cal U}_{\rm q}$ for the MW (orange), OGLE-II LMC (blue), OGLE-III LMC  (green), and OGLE-II SMC (red).  We also display ${\cal P}_{\rm deep}$ and ${\cal P}_{\rm med}$ determined by incorporating the low-metallicity $Z$ = 0.004 stellar tracks and assuming a uniform mass-ratio distribution (black).  The probabilities ${\cal P}_{\rm med}$ based on the \citet{Soderhjelm2005} solar-metallicity results (magenta) are consistent with our MW values.  The probabilities of observing eclipses decreases with increasing $P$ due to geometrical selection effects, and also decreases with mass-ratio distributions which favor low-mass, smaller companions.}
\end{figure}

\subsubsection{Intrinsic Period Distribution ${\cal U}_{\rm P}$}

We now fit the observed eclipsing binary period distributions ${\cal O}_{\rm deep}(P)$ or ${\cal O}_{\rm med}(P)$ only, which constrain the intrinsic period distributions ${\cal U}_{\rm P}$ and the normalizations to ${\cal F}_{\rm close}$ according to Eq.~1.    We minimize the $\chi^2_{\rm P}$($\vec{x}$) statistics between the measured eclipsing binary period distributions ${\cal O}_{\rm deep}$(log\,$P$)  and our Monte Carlo models ${\cal M}_{\rm deep}$(log\,$P$, $\vec{x}$):

\begin{equation}
 \chi^2_{\rm P}(\vec{x}) = \sum_k^{N_{\rm P}} \Big(\frac{{\cal O}_{\rm deep}({\rm log}\,P_k) - {\cal M}_{\rm deep}({\rm log}\,P_k, \vec{x})}{\mbox{\Large $\sigma$}_{{\cal O}_{\rm deep}}({\rm log}\,P_k)}\Big)^2
\end{equation}

\noindent We calculate similar statistics for the medium eclipse depth samples.  We sum over the logarithmic period bins of data displayed in Figure 9, specifically the $N_{\rm P}$ = 10 bins of ${\cal O}_{\rm deep}(P)$ for the OGLE-II LMC and SMC populations, $N_{\rm P}$ = 3 bins of ${\cal O}_{\rm med}(P)$ for the MW,  and the $N_{\rm P}$ = 10 bins of ${\cal O}_{\rm med}(P)$  for the OGLE-III LMC sample.  The measured period distribution constrains $\gamma_{\rm P}$ and ${\cal F}_{\rm close}$, which effectively gives $\nu$ =  $N_{\rm P}$\,$-$\,2 degrees of freedom. As in \S3.3.1, we report the $\chi^2_{\rm P}$ statistics and fitted model parameters in Table 3 as well as display the two-dimensional probability contour of ${\cal F}_{\rm close}$ versus $\gamma_{\rm P}$ in Figure 10.  

\begin{figure}[H]
\center
\includegraphics[width=7.2in]{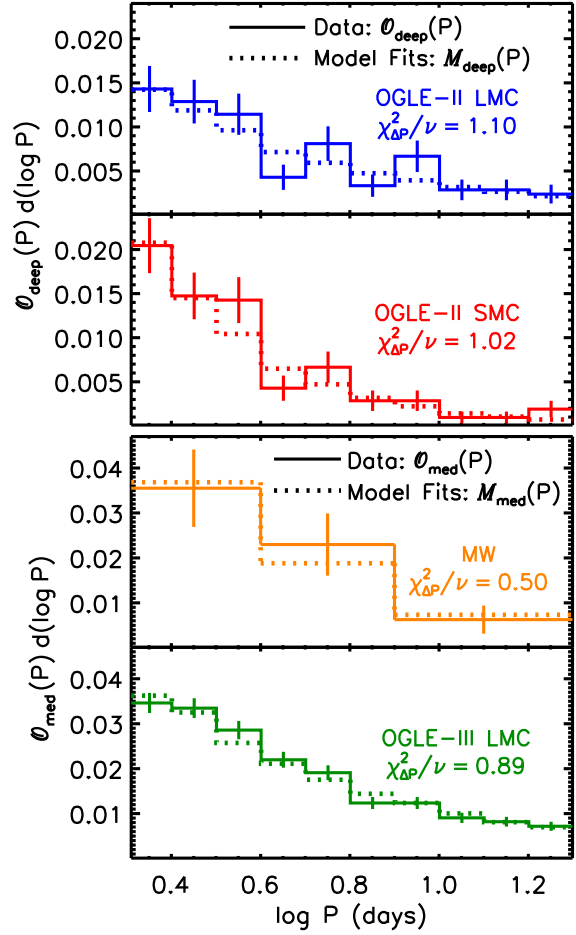}
\caption{The observed eclipsing binary period distributions (solid) for deep eclipses ${\cal O}_{\rm deep}(P)$ (top two panels) and extension toward medium eclipse depths ${\cal O}_{\rm med}(P)$ (bottom two panels) as displayed in Figure 2 for the Hipparcos MW (orange), OGLE-II LMC (blue), OGLE-III LMC (green), and OGLE-II SMC (red) populations.  We determine the best-fit Monte Carlo models ${\cal M}_{\rm deep}(P)$ and ${\cal M}_{\rm med}(P)$ (dotted) by minimizing the $\chi^2_{\rm P}$ statistic across the logarithmic period bins of data.}
\end{figure}

\begin{table}[H]
\small
\center
\begin{tabular}{|c|c|c|c|c|c|c|}
\hline
Sample       & Eclipse Depths & $\chi^2_{\rm P}/{\nu}$ & $\nu$ & PTE  & $\gamma_{\rm P}$   & ${\cal F}_{\rm close}$  \\
\hline
MW           &   Medium \& Deep      & 0.50                   &   1   & 0.48 & $-$0.4\,$\pm$\,0.3 & 0.22\,$\pm$\,0.06       \\
\hline
OGLE-II LMC  &   Deep         & 1.10                   &   8   & 0.36 & $-$0.3\,$\pm$\,0.2 & 0.22\,$\pm$\,0.08       \\
\hline
OGLE-III LMC &   Medium \& Deep      & 0.89                   &   8   & 0.53 & $-$0.1\,$\pm$\,0.2 & 0.24\,$\pm$\,0.05       \\ 
\hline
OGLE-II SMC  &   Deep         & 1.02                   &   8   & 0.42 & $-$0.9\,$\pm$\,0.2 & 0.21\,$\pm$\,0.09       \\
\hline
\end{tabular}
\caption{Results of our Monte Carlo simulations and fits to the observed eclipsing binary period distributions ${\cal O}_{\rm deep}(P)$ or ${\cal O}_{\rm med}(P)$ {\it only}.  For each of the eclipsing binary samples, we list whether the deep  eclipse ${\cal O}_{\rm deep}(P)$ or extension toward medium eclipse depth ${\cal O}_{\rm med }(P)$ samples were used to fit the period distribution, minimized reduced $\chi^2_{\rm P}$ statistics, degrees of freedom $\nu$ = $N_{\rm P}$ $-$ 2, probabilities to exceed $\chi^2_{\rm P}$ given~$\nu$, and the mean values and 1$\sigma$ uncertainties of the two model parameters constrained by ${\cal O}_{\rm deep}(P)$ or ${\cal O}_{\rm med}(P)$.  }
\end{table}

\begin{figure}[H]
\center
\includegraphics[width=6.7in, trim= 0cm 3.6cm 0cm 4.1cm,clip=true]{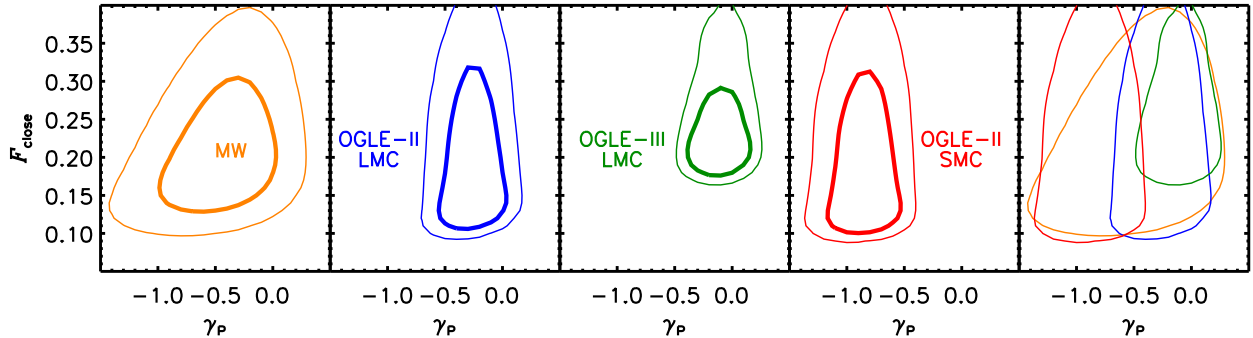}
\caption{Probability contours at the 1$\sigma$ (thick) and 2$\sigma$ (thin) confidence levels of ${\cal F}_{\rm close}$ versus $\gamma_{\rm P}$ constrained {\it only} by the observed eclipse depth distributions ${\cal O}_{\rm deep}(P)$ or ${\cal O}_{\rm med}(P)$ for  the Hipparcos MW (orange), OGLE-II LMC (blue), OGLE-III LMC (green), and OGLE-II SMC (red) populations.  Although the OGLE-II SMC population favors a distribution that is skewed toward shorter periods while the OGLE-III LMC population is consistent with \"{O}pik's law of $\gamma_{\rm P}$ = 0, all four samples are mildly consistent with ${\cal F}_{\rm close}$ $\approx$ 20\% and $\gamma_{\rm P}$~$\approx$~$-$0.4.   }
\end{figure}

By making simple approximations in \S2, we showed that all four eclipsing binary samples were skewed toward shorter periods relative to \"{O}pik's prediction of ${\cal S}_{\rm deep}(P)$\,d(log\,$P$) $\propto$ ${\cal S}_{\rm med}(P)$\,d(log\,$P$) $\propto$ $P^{-2/3}$\,d(log\,$P$). We confirm this result with our more robust light curve modeling and Monte Carlo simulations, where we find fitted mean values of $\gamma_{\rm P}$ that are negative for all four main samples.  However, the OGLE-III LMC value of $\gamma_{\rm P}$ = $-$0.1\,$\pm$\,0.2 is still consistent with \"{O}pik's law of $\gamma_{\rm P}$~=~0, while the OGLE-II SMC population is significantly skewed toward shorter periods with $\gamma_{\rm P}$~=~$-$0.9\,$\pm$\,0.2.  These two values for $\gamma_{\rm P}$ are discrepant at the 2.4$\sigma$ level.  This is similar to our K-S test in \S2 between the OGLE-II SMC and OGLE-III LMC unbinned ${\cal O}_{\rm deep}(P)$ data, which gave a probability of consistency of $p_{\rm KS}$ = 0.01. 

As discussed in \S2, it is possible that long period systems $P$ $>$ 10 days with moderate eclipse depths $\Delta$m = 0.25\,-\,0.30 mag have remained undetected in the OGLE-II SMC sample because their members are systematically 0.5 mag fainter.    If we only use the OGLE-II SMC data with $P$~=~2$~\mbox{-}~$10~days and  $\Delta$m = 0.30\,-\,0.65 mag to constrain our fit, then we find $\gamma_{\rm P}$ = $-$0.7 $\pm$ 0.4, which is more consistent with the LMC result.  In any case, whether the slight discrepancy is intrinsic or due to small systematics, the best-fitting period exponent for the MW of $\gamma_{\rm P}$ $\approx$ $-$0.4 is between the LMC and SMC values.  We confirm this intermediate value based on spectroscopic radial velocity observations of nearby early-type stars (see \S4).  Although there is a strong indication that the SMC period distribution is skewed  toward shorter periods compared to the LMC data, there is no clear trend with metallicity.  Moreover, the MW, SMC and LMC samples are all mildly consistent, i.e. less than 2$\sigma$ discrepancy,  with the intermediate value of $\gamma_{\rm P}$~$\approx$~$-$0.4.

\subsubsection{Close Binary Fraction ${\cal F}_{\rm close}$}

The close binary fractions ${\cal F}_{\rm close}$  are not well constrained by fitting the observed eclipse depth and  period distributions separately.  For example, the 1$\sigma$ errors in the close binary fractions from only fitting ${\cal O}_{\Delta {\rm m}}$ were $\delta {\cal F}_{\rm close}$ $\approx$ 0.05\,-\,0.08, depending on the sample (see Table 2), while the errors from only fitting ${\cal O}_{\rm deep}(P)$ or  ${\cal O}_{\rm med}(P)$ were $\delta {\cal F}_{\rm close}$ $\approx$ 0.05\,-\,0.09 (Table 3).  To measure ${\cal F}_{\rm close}$ most precisely, we now fit ${\cal O}_{\Delta {\rm m}}$ and either ${\cal O}_{\rm deep}(P)$ or  ${\cal O}_{\rm med}(P)$ simultaneously by minimizing $\chi^2$~=~$\chi^2_{\Delta {\rm m}}$~+~$\chi^2_{\rm P}$.  For each sample, we sum over the same bins of eclipse depths and orbital periods that are complete as reported in \S3.3.1 and \S3.3.3, respectively.   This combined fit gives $\nu$~=~$N_{\Delta {\rm m}}$~+~$N_{\rm P}$~$-$~4 degrees of freedom since all four model parameters are constrained. In Table~4, we report the fitting statistics as well as the  means and 1$\sigma$ uncertainties for ${\cal F}_{\rm close}$ only because this combined method does not alter our previous estimates of $\gamma_{\rm q}$, ${\cal F}_{\rm twin}$, and $\gamma_{\rm P}$.   The $\chi^2/\nu$ values are all close to unity and the probabilities to exceed are in the 1$\sigma$ range 0.16$\,\mbox{-}\,$0.84, demonstrating our models are sufficient in explaining the data.

\begin{table}[H]
\small
\center
\begin{tabular}{|c|c|c|c|c|c|}
\hline
Sample       & Eclipse Depths &   $\chi^2/{\nu}$ & $\nu$ & PTE  & ${\cal F}_{\rm close}$ \\
\hline
MW           &  Medium \& Deep      &      0.44       &   4   & 0.76 &   0.22\,$\pm$\,0.04 \\
\hline
OGLE-II LMC  &  Deep      &      0.89       &  14   & 0.58 &   0.21\,$\pm$\,0.06 \\
\hline
OGLE-III LMC &  Medium \& Deep      &      1.02       &  17   & 0.39 &   0.28\,$\pm$\,0.02 \\ 
\hline
OGLE-II SMC  &  Deep      &      0.81       &  14   & 0.68 &   0.23\,$\pm$\,0.06 \\
\hline
\end{tabular}
\caption{Results of our fits to the observed eclipse depth distributions ${\cal O}_{\Delta {\rm m}}$ {\it and}  observed eclipsing binary period distributions  ${\cal O}_{\rm deep}(P)$ or  ${\cal O}_{\rm med}(P)$. For each sample, we list whether the deep or extension toward medium eclipse depth samples were used to simultaneously fit the eclipse depth and period distributions.  We also report the minimized reduced $\chi^2$ = $\chi^2_{\Delta {\rm m}}$ + $\chi^2_{\rm P}$ statistics, degrees of freedom $\nu$ = $N_{\Delta {\rm m}}$ + $N_{\rm P}$ $-$ 4, probabilities to exceed $\chi^2$ given $\nu$, and the mean values and 1$\sigma$ uncertainties of the close binary fractions ${\cal F}_{\rm close}$ {\it before correcting for Malmquist bias and propagating systematic errors}. }
\end{table}

In order to fit ${\cal O}_{\Delta {\rm m}}$ and either ${\cal O}_{\rm deep}(P)$ or  ${\cal O}_{\rm med}(P)$ simultaneously, we have assumed $\Delta$m and $P$ are independent so that $p$ $\propto$ $e^{-\chi^2_{\Delta {\rm m}}/2}$\,$\times$\,$e^{-\chi^2_{\rm P}/2}$ = $e^{-(\chi^2_{\Delta {\rm m}}+\chi^2_{\rm P})/2}$ = $e^{-\chi^2/2}$.   For all four samples of eclipsing binaries, the Spearman rank correlation coefficients between $\Delta$m and $P$ are rather small at  $|\rho|$ $<$ 0.15 across the eclipse depth intervals which are complete. These small coefficients justify our procedure for fitting the eclipsing binary period and eclipse depth distributions together in order to better constrain ${\cal F}_{\rm close}$.  Moreover, the probability of observing medium eclipses ${\cal P}_{\rm med}$($P$) determined in \S3.3.2 only marginally depends on the underlying mass-ratio distribution ${\cal U}_{\rm q}$.  Therefore, any trend between mass-ratios and orbital periods will not affect the fitted close binary fractions beyond the quantified errors.  

If we had used simple prescriptions for eclipse depths instead of the detailed \textsc{nightfall} light curve models, our fitted values for ${\cal F}_{\rm close}$ would have been a factor of (10\,-\,20)\% different, i.e. $\Delta{\cal F}_{\rm close}$~$\approx$~0.02$\,\mbox{-}\,$0.04 depending on the sample with no general trend on the direction.  This would have been a dominant source of  error, especially for the OGLE-III LMC data, so it was imperative that we implemented the more precise \textsc{nightfall} simulations.  Before we comment further on our measurements of ${\cal F}_{\rm close}$ in the different environments, we must first correct for Malmquist bias.

\subsection{Malmquist Bias}

\subsubsection{Milky Way}

 Unresolved binaries, including eclipsing systems, are systematically brighter than their single star counterparts.  For a magnitude-limited sample within our MW,  more luminous binaries are probed over a larger volume than their single star counterparts, which causes the binary fraction to be artificially enhanced.  This classical Malmquist bias is sometimes referred to as the  \citet{Opik1923} or \citet{Branch1976} effect in the context of binary stars.  

Of the ${\cal N}_{\rm med}$ = 31 eclipsing binaries in our medium eclipse depth MW sample with $\langle$H$_{\rm P}\rangle$~$<$~9.3, only four systems are fainter than $\langle$H$_{\rm P}\rangle$~$>$~8.8 \citep{Lefevre2009}.  One of these systems, V2126~Cyg, has a moderate magnitude of $\langle$H$_{\rm P}\rangle$ = 9.0 and shallow eclipse depth of $\Delta$H$_{\rm P}$ = 0.13.  This small eclipse depth indicates a faint, low-mass companion, although the less likely scenario of a grazing eclipse with a more massive secondary is also feasible.  The remaining three systems, IT~Lib, LN~Mus, and TU~Mon, all have fainter system magnitudes $\langle$H$_{\rm P}\rangle$~$>$~9.1 and deeper eclipses $\Delta$H$_{\rm P}$~$>$~0.18, suggesting that their primaries alone do not fall within our magnitude limit of $\langle$H$_{\rm P}\rangle$~$<$~9.3.  If we remove this excess number of ${\cal N}_{\rm ex}$ = 3\,-\,4 eclipsing binaries from both our eclipsing binary sample ${\cal N}_{\rm med}$  as well as from the total number of systems ${\cal N}_{\rm B}$, then the eclipsing binary fraction with medium eclipse depths ${\cal F}_{\rm med}$ = ${\cal N}_{\rm med}$/${\cal N}_{\rm B}$ would decrease by a factor of $\approx$11\%, i.e. $\Delta{\cal F}_{\rm med}$ $\approx$ $-$0.002.  

However, we must also remove from the denominator ${\cal N}_{\rm B}$ {\it other} binaries with luminous secondaries which have primaries that fall below our magnitude limit.  These include close binaries that remain undetected because they have orientations which do not produce observable eclipses.  Based on the correction factor ${\cal C}_{\rm med,i}$ = 9\,$\pm$\,2 for geometrical selection effects alone for the MW sample (see \S3.3.2), then we expect a total of ${\cal N}_{\rm med}$\,$\times$\,${\cal C}_{\rm med,i}$ $\approx$ 30 binaries with $P$ = 2\,-\,20 days that should be removed from ${\cal N}_{\rm B}$.  

Additional systems that contaminate ${\cal N}_{\rm B}$ consist of binaries with luminous secondaries outside of our period range of $P$ = 2\,-\,20 days. To estimate their contribution toward Malmquist bias, we calculate the ratio ${\cal R}_{\rm P}$ between the frequency of massive secondaries across all orbital periods  to the frequency of massive secondaries with $P$ = 2\,-\,20 days.  Spectroscopic observations of O and B type stars in the MW reveal 0.16\,-\,0.31 companions with $q$ $>$ 0.1 per decade of orbital period  at log~$P$~$\approx$~0.8 \citep[][see also \S4]{Garmany1980,Levato1987,Abt1990,Sana2012}.  At longer orbital periods of log $P$ $\approx$ 6.5, photometric observations of visually resolved binaries give a lower value of  $\approx$ 0.10\,-\,0.16 companions with $q$ $\gtrsim$ 0.1 per decade of orbital period \citep{Duchene2001,Shatsky2002,Turner2008,Mason2009}. Using these two points to anchor the slope of the  period distribution, we integrate from log $P$ = 0.1 to the widest, stable orbits of log $P$ $\approx$ 8.5.  We find there are 6.4\,$\pm$\,1.3   as many total companions as there are binaries with $P$~=~2$\,\mbox{-}\,$20~days.  However, longer period binaries with $P$ $>$ 20 days may have a mass-ratio distribution that differs from our sample at shorter orbital periods.  For example,  \citet{Abt1990} and \citet{Duchene2001} suggest random pairings of the initial mass function for wide binaries so that $\gamma_{\rm q}$ $\approx$ $-$2.3,  the distribution of \citet{Preibisch1999} indicates a more moderate value of $\gamma_{\rm q}$ $\approx$ $-$1.5, while \citet{Shatsky2002} gives $\gamma_{\rm q}$ $\approx$ $-$0.5 for visually resolved binaries, which is consistent with the values inferred from our close eclipsing binary samples of $\gamma_{\rm q}$~$\approx$~$-$1.0$\,\mbox{-}\,$$-0.6$.   Assuming $\gamma_{\rm q}$ = $-$1.5\,$\pm$\,0.5 for binaries outside our period range, then there are 2.3\,$\pm$\,1.1 times fewer binaries with $q$ $>$ 0.6 relative to the mass-ratio distribution constrained for our close eclipsing binaries.  Since we are primarily concerned with massive secondaries which contribute toward Malmquist bias, then ${\cal R}_{\rm P}$ $\approx$ (6.4\,$\pm$\,1.3)/(2.3\,$\pm$\,1.1) = 2.8\,$\pm$\,1.4.

The eclipsing binary fraction for the MW sample after correcting for classical Malmquist bias is then:

\begin{equation}
 {\cal F}_{\rm med} = \frac{{\cal N}_{\rm med} - {\cal N}_{\rm ex}}{{\cal N}_{\rm B} - {\cal N}_{\rm ex} {\cal C}_{\rm med, i} {\cal R}_{\rm P}} = (1.83\,\pm\,0.38)\%
\end{equation}

\noindent where we propagated the uncertainties in ${\cal C}_{\rm med, i}$ and ${\cal R}_{\rm P}$ as well as the Poisson errors in ${\cal N}_{\rm med}$ and ${\cal N}_{\rm ex}$. Note that removing non-eclipsing binaries with luminous secondaries that remain undetected  mitigates the effects of Malmquist bias.  Specifically, we find the reduction factor to be ${\cal C}_{\rm Malm}$~=~0.94\,$\pm$\,0.05 instead of the factor  of ${\cal C}_{\rm Malm}$ = 0.89 determined above when we only removed ${\cal N}_{\rm ex}$ eclipsing systems.   Although these two competing effects in the numerator and denominator of the above relation have been discussed in the literature \citep[e.g.][]{Bouy2003}, the removal of binaries with luminous secondaries which remain undetected is typically neglected.  The inferred close binary fraction for the MW will also decrease by a factor of ${\cal C}_{\rm Malm}$ $=$ 0.94, so that the corrected value is only slightly lower at ${\cal F}_{\rm close}$ = 21\% (see \S3.5).  

\subsubsection{Magellanic Clouds}

  In the case of the Magellanic Clouds at fixed, known distances, classical Malmquist bias does not apply.  Nonetheless, our absolute magnitude interval  of $\overline{\rm M}_{\rm I}$ = [$-$3.8,\,$-$1.3] contain binaries with primaries which are lower in intrinsic luminosity and stellar mass relative to single stars in the same magnitude range.  Some binaries in our sample have primaries that are fainter than our magnitude limit of M$_{\rm I}$ = $-$1.3, while some systems have primaries in the range we want to consider but are pushed beyond M$_{\rm I}$ = $-$3.8 because of the excess light added by the secondary.  Since the number of primaries dramatically increases with decreasing stellar mass and luminosity, the net effect is that the binary fractions are biased toward larger values.  Hence, our statistics are affected by Malmquist bias of the second kind because two classes of objects, e.g. binaries and single stars, are surveyed to a certain depth down their respective luminosity functions \citep{Teerikorpi1997,Butkevich2005}. 

  For example, \citet{Mazeh2006} used OGLE-II data of the LMC to identify 938 eclipsing binaries on the MS with apparent magnitudes 17 $<$ I $<$ 19 and periods 2 $<$ $P$ (days) $<$ 10.  Instead of normalizing these eclipsing binaries to the total number of $\approx$\,330,000 MS systems with 17~$<$~I~$<$~19, they assumed the average eclipsing binary was $\langle \Delta$M$_{\rm I} \rangle$ = 0.5 mag brighter than the primary component alone, and therefore normalized to the $\approx$\,700,000 MS systems with 17.5~$<$~I~$<$~19.5.   Their correction for Malmquist bias of the second kind lowered the inferred close binary fraction by a factor of 2.1, i.e. ${\cal C}_{\rm Malm}$ = 0.48.  

  Instead of adding systems below our lower magnitude limit as done by \citet{Mazeh2006}, we remove binaries with luminous secondaries within our magnitude interval $\overline{\rm M}_{\rm I}$ = [$-$3.8,\,$-$1.3] as described above for the MW.  To determine the average fraction $\langle \delta {\cal F}_{\rm I} \rangle$ of eclipsing binaries that should be removed from our Magellanic Cloud samples, we use the OGLE photometric catalogs \citep{Udalski1998,Udalski2000,Udalski2008} to compute the observed fractional decrease $\delta{\cal F}_{\rm I}$ in the total number of MS systems as a function of incremental I-band magnitude $\Delta$M$_{\rm I}$.  Quantitatively:

\begin{equation}
 \delta{\cal F}_{\rm I} (\Delta{\rm M}_{\rm I}) = 1 - \frac{{\cal N}(\overline{\rm M}_{\rm I} - \Delta{\rm M}_{\rm I})}{{\cal N}(\overline{\rm M}_{\rm I})} 
\end{equation}

\noindent where ${\cal N}(\overline{\rm M}_{\rm I})$ = ${\cal N}_{\rm B}$ is our original total number of MS systems and ${\cal N}(\overline{\rm M}_{\rm I} - \Delta{\rm M}_{\rm I})$ is the number of systems with colors V$-$I $<$ 0.1 in the interval M$_{\rm I}$ = [$-$3.8,\,$-$1.3\,$-$\,$\Delta$M$_{\rm I}$].  We display $\delta{\cal F}_{\rm I}$ in the top panel of Figure 11 for the three OGLE Magellanic Cloud samples.  We only show the fractional decreases $\delta{\cal F}_{\rm I}$ across the interval 0 $<$ $\Delta$M$_{\rm I}$ $<$ 0.75 because binary companions can only contribute a luminosity excess in this range.  The three distributions of  $\delta{\cal F}_{\rm I}$ are similar among the three populations due to the consistency of the stellar mass function in the different environments.  The total number of systems is approximately halved, i.e. $\delta{\cal F}_{\rm I}$ = 0.5, at $\Delta$M$_{\rm I}$ $\approx$ 0.5, consistent with the result of \citet{Mazeh2006}.

Instead of assuming an average value for the magnitude difference $\langle\Delta$M$_{\rm I}\rangle$ = 0.5 mag between a single star and eclipsing binary with the same primary, we use the OGLE eclipsing binary data and our Monte Carlo simulations to model an I-band excess probability distribution $p_{\rm I}$\,($\Delta$M$_{\rm I}$)\,d($\Delta$M$_{\rm I}$). Using the best-fit models for each of the three OGLE samples, we synthesize distributions of secondary masses which produce observable eclipses, i.e. systems with eclipse depths 0.25~$<$~$\Delta$m~$<$~0.65 for our deep samples and 0.1~$<$~$\Delta$m~$<$~0.65 for our extension toward medium eclipse depths (OGLE-III LMC only).  We then use the stellar tracks of \citet{Bertelli2009} as well as color indices and bolometric corrections of \citet{Cox2000} to convert the distribution of secondary masses that produce observable eclipses into a distribution of secondary absolute magnitudes in the I-band.  We can then easily determine the system luminosity, the luminosity of the primary alone, and the I-band excess $\Delta$M$_{\rm I}$ between the two for each eclipsing binary.  In the bottom panel of Figure 11, we display our results for the I-band excess probability distribution $p_{\rm I}$\,($\Delta$M$_{\rm I}$)\,d($\Delta$M$_{\rm I}$), which is normalized so that the distribution integrates to unity.

\begin{figure}[H]
\center
\includegraphics[width=6.5in]{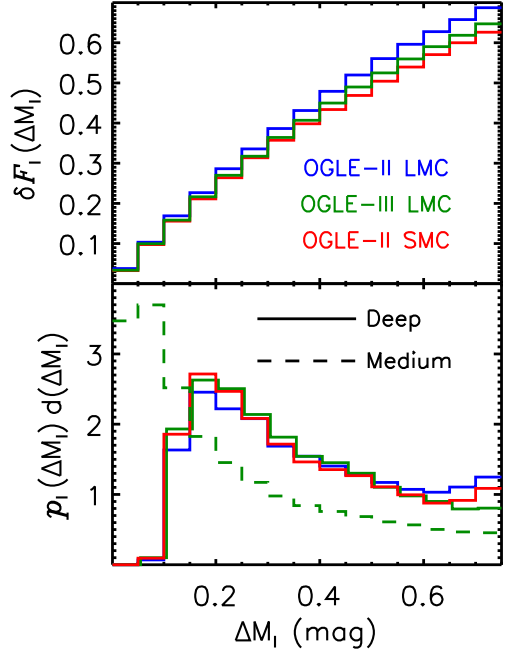}
\caption{Top panel: the observed fractional decrease $\delta{\cal F}_{\rm I}$ in the total number of MS systems as a function of incremental I-band magnitude $\Delta$M$_{\rm I}$ for the OGLE-II LMC (blue), OGLE-III LMC (green), and OGLE-II SMC (red) samples. Bottom panel: based on our best-fit Monte Carlo simulations, the modeled I-band excess probability distributions $p_{\rm I}$\,($\Delta$M$_{\rm I}$)\,d($\Delta$M$_{\rm I}$) of binaries exhibiting deep (solid) and medium (dashed) eclipses due to increased luminosity from the companion.  In order to correct for Malmquist bias of the second kind, we determine the average fraction $\langle \delta{\cal F}_{\rm I}\rangle$ of eclipsing binaries that should be removed from our samples according to $\langle \delta{\cal F}_{\rm I}\rangle$ = $\int \delta{\cal F}_{\rm I}$($\Delta$M$_{\rm I}$)\,$p_{\rm I}$($\Delta$M$_{\rm I}$)\,d($\Delta$M$_{\rm I}$). }
\end{figure}

The I-band excess probability distributions $p_{\rm I}$ for the three OGLE samples exhibiting deep eclipses are all quite similar.  This is because they have similar eclipse depth distributions ${\cal O}_{\Delta {\rm m}}$, and therefore similar mass-ratio distributions ${\cal U}_{\rm q}$.    Very few low-mass, low-luminosity secondaries with $\Delta$M$_{\rm I}$~$<$~0.1 mag are capable of producing deep eclipses with 0.25 $<$ $\Delta$m $<$ 0.65. However, many of these faint secondaries are included in the OGLE-III LMC medium eclipse depth sample.  The median I-band excess is only $\langle \Delta$M$_{\rm I} \rangle$ = 0.35 and $\langle \Delta$M$_{\rm I} \rangle$ = 0.20 mag for the deep and medium samples, respectively, which are lower than the value of $\langle \Delta$M$_{\rm I} \rangle$ = 0.5 used by \citet{Mazeh2006}.  Note that these values of $\langle \Delta$M$ \rangle$ = 0.2\,-\,0.5 mag are the reason we excluded the ${\cal N}_{\rm ex}$ = 3\,-\,4 eclipsing binaries in the MW sample (\S3.4.1) that were within 0.2\,-\,0.5 mag of our magnitude limit of $\langle$H$_{\rm P}\rangle$~=~9.3.  

 We can now compute the average fraction $\langle\delta{\cal F}_{\rm I}\rangle$ of eclipsing binaries that should be removed from our samples by weighting $\delta{\cal F}_{\rm I}$ with the I-band excess probability distribution, i.e. $\langle\delta{\cal F}_{\rm I}\rangle$ = $\int \delta{\cal F}_{\rm I}$($\Delta$M$_{\rm I}$)\,$p_{\rm I}$($\Delta$M$_{\rm I}$)\,d($\Delta$M$_{\rm I}$).   We find $\langle \delta{\cal F}_{\rm I} \rangle$ = 0.38\,$\pm$\,0.11 and $\langle \delta{\cal F}_{\rm I} \rangle$ = 0.35\,$\pm$\,0.10 for the OGLE-II LMC and SMC deep eclipse samples, respectively, and $\langle \delta{\cal F}_{\rm I} \rangle$ = 0.23\,$\pm$\,0.08 for the OGLE-III LMC medium eclipse sample.  These values are lower than the estimate of $\langle \delta{\cal F}_{\rm I} \rangle$ = 0.52 by \citet{Mazeh2006} because the modeled I-band excess probability distributions are weighted more toward fainter companions.  

Instead of only removing this average fraction $\langle\delta{\cal F}_{\rm I}\rangle$ of eclipsing binaries, i.e. assuming ${\cal C}_{\rm Malm}$ = 1\,$-$\,$\langle \delta{\cal F}_{\rm I} \rangle$, we must also account for the other binaries with luminous secondaries outside our parameter space of eclipse depths and orbital periods.  Using a similar format as in Eq. 5, we derive:

\begin{equation}
 {\cal C}_{\rm Malm} = \frac{1 - \langle \delta{\cal F}_{\rm I} \rangle}{1 - {\cal F}_{\rm med} \langle \delta{\cal F}_{\rm I} \rangle {\cal C}_{\rm med, i} {\cal R}_{\rm P}}
\end{equation}

\noindent where ${\cal F}_{\rm med}$ = 1.87\% is the uncorrected eclipsing binary fraction in Table 1 and ${\cal C}_{\rm med,i}$ = 11\,$\pm$\,2 is the correction factor for geometrical selection effects alone (see \S3.3.2) for the OGLE-III LMC medium sample, and ${\cal R}_{\rm P}$ = 2.8\,$\pm$\,1.4 has the same definition as in \S3.4.1.  We calculate similar values for the OGLE-II LMC and SMC deep eclipse samples, where ${\cal F}_{\rm deep}$ = 0.70\% and ${\cal C}_{\rm deep,i}$~=~14\,$\pm$\,3.  We find the overall correction factors for Malmquist bias of the second kind to be ${\cal C}_{\rm Malm}$ = 0.73\,$\pm$\,0.16, 0.91\,$\pm$\,0.12, and 0.76\,$\pm$\,0.15  for the OGLE-II LMC, OGLE-III LMC, and OGLE-II SMC samples respectively.  Because the OGLE-III LMC survey was sensitive to shallow eclipses that systematically favored low-luminosity companions with $\langle \Delta$M$_{\rm I}\rangle$ $\approx$ 0.2 mag, the correction for Malmquist bias for this population is nearly negligible.

\subsection{Corrected Results}

We have implemented detailed \textsc{nightfall} light curve models (\S3.1) and computed thousands of Monte Carlo simulations (\S3.2) in order to correct for geometrical selection effects and incompleteness toward low-mass companions.  By fitting the observed eclipsing binary distributions using various methods, we have derived the underlying intrinsic binary properties for the MW, LMC, and SMC (\S3.3).  Because our eclipsing binary samples are magnitude limited and therefore  subject to Malmquist bias, we have determined accurate reduction factors (\S3.4) by incorporating the observed stellar luminosity functions, modeling the I-band excess probability distributions, and accounting for other binaries outside our parameter space of eclipsing systems.  We have also quantified many sources of systematic errors in our analysis, including the single-mass primary approximation (factor of 8\% uncertainty for the MW and 10\% for the Magllanic Cloud samples, i.e. $\delta {\cal F}_{\rm close}$ $\approx$ 0.02), the contribution of the few giants and evolved primaries filling their Roche lobes (factor of 3\%), the conversion of Roche-lobe filling factors (factor of 7\%), effects of eccentric orbits (factor of 2\%), third light contamination due to triple systems and stellar blending (factor of 6\%), and the uncertainties in the Malmquist bias reduction factors (factors of 5\,-\,16\%, depending on the sample). Assuming Gaussian uncertainties, we add these systematic errors in quadrature and propagate the total factor of (14-21)\% systematic uncertainty,  i.e. $\delta {\cal F}_{\rm close}$ $\approx$ 0.03\,-\,0.04 depending on the sample, into our evaluations of the close binary fraction.     

Based on our $\chi^2$ fits, correction for Malmquist bias, and propagation of systematic errors, our finalized results for ${\cal F}_{\rm close}$ are 0.21\,$\pm$\,0.05, 0.16\,$\pm$\,0.06, 0.25\,$\pm$\,0.04, and 0.17\,$\pm$\,0.06 for the MW, OGLE-II LMC, OGLE-III LMC, and OGLE-II SMC populations, respectively.  We list these corrected values in Table 5.  All of the close binary fractions  ${\cal F}_{\rm close}$ are consistent with each other at the 1.2$\sigma$ level. The fact that all four environments have ${\cal F}_{\rm close}$ = (16\,-\,25)\% demonstrates that the close binary fraction does not substantially vary across metallicities log($Z$/\Zsun) $\approx$ $-$0.7 - 0.0.

\begin{table}[H]
\small
\center
\begin{tabular}{|c|c|c|c|c|}
\hline
                          & MW & OGLE-II LMC & OGLE-III LMC & OGLE-II SMC \\
\hline
${\cal F}_{\rm close}$    & (21\,$\pm$\,5)\% & (16\,$\pm$\,6)\% & (25\,$\pm$\,4)\% & (17\,$\pm$\,6)\% \\
\hline
\end{tabular}
\caption{For the four different eclipsing binary samples, we list the corrected fractions of early-B stars with companions $q$ $>$ 0.1 at orbital periods $P$ = 2\,-\,20 days after accounting for geometrical selection effects, incompleteness toward low-mass companions, Malmquist bias, and systematic errors. }
\end{table}

  Instead of inferring the intrinsic period distributions ${\cal U}_{\rm P}$ from our fitted model parameters $\gamma_{\rm P}$ and ${\cal F}_{\rm close}$, we can also visualize the distributions based on the observed eclipsing binary period distributions (see \S2) and our modeled probabilities of observing eclipses  (see \S3.3.2).  For the OGLE-II LMC and SMC samples, we use ${\cal U}_{\rm P}(P)$\,d(log\,$P$) = [${\cal O}_{\rm deep}(P)$\,d(log\,$P$)\,/\,${\cal P}_{\rm deep}(P)$]$\times$$C_{\rm Malm}$, where $C_{\rm Malm}$ $\approx$ 0.75 is the slight correction factor for Malmquist bias (\S3.4). Similarly, we  use ${\cal U}_{\rm P}(P)$\,d(log\,$P$) = [${\cal O}_{\rm med}(P)$\,d(log\,$P$)\,/\,${\cal P}_{\rm med}(P)$]$\times$$C_{\rm Malm}$, where $C_{\rm Malm}$ = 0.91 for the OGLE-III LMC population and $C_{\rm Malm}$ = 0.94  for the MW.  We present the results in Figure~12, where we have propagated in quadrature the errors from each of the three terms in the relations for ${\cal U}_{\rm P}(P)$.  

 At short periods $P$ = 2\,-\,4 days, the populations have ${\cal U}_{\rm P}$ $\approx$ 0.2\,-\,0.3 companions with $q$~$>$~0.1 per full decade of period.  At longer periods $P$ = 10\,-\,20 days, the values are slightly lower at ${\cal U}_{\rm P}$~$\approx$~0.1$\,\mbox{-}\,$0.2.   Even after correcting for geometrical selection effects and incompleteness toward low-mass companions, the general trend is that ${\cal U}_{\rm P}$ decreases with increasing $P$ across the interval 0.3~$<$~log\,$P$~$<$~1.3.  This is consistent with our $\chi^2_{\rm P}$ fits which favored negative $\gamma_{\rm P}$, i.e. distributions skewed toward shorter periods compared to \"{O}pik's law of $\gamma_{\rm P}$ = 0.  The integrated fractions cover a narrow range ${\cal F}_{\rm close}$ = $\int {\cal U}_{\rm P}$\,d(log\,$P$) = 0.16\,-\,0.25, again demonstrating the close binary fraction does not change with metallicity.  

\begin{figure}[H]
\center
\includegraphics[width=4.0in]{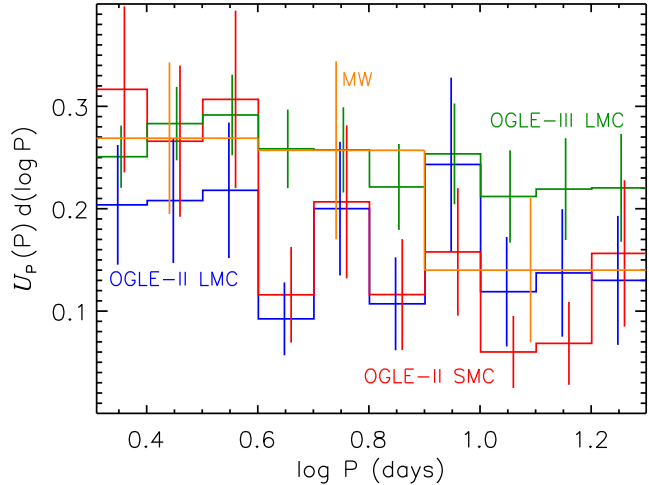}
\caption{The corrected intrinsic period distribution ${\cal U}_{\rm P}$, i.e. the frequency of companions with $q$ $>$ 0.1 per full decade of period, for the MW (orange), OGLE-II LMC (blue), OGLE-III LMC (green), and OGLE-II SMC (red) populations. All the distributions favor a period distribution that decreases slightly with increasing period, even after correcting for geometrical selection effects.   The small range in the integrated fractions ${\cal F}_{\rm close}$ = $\int {\cal U}_{\rm P}$\,d(log\,$P$) = (16\,-\,25)\% attests to the uniformity of the early-B close binary fraction.  }
\end{figure}

\section{Comparison to Spectroscopic Binaries in the MW}

We have utilized the \citet{Lefevre2009} catalog of eclipsing binaries based on Hipparcos data to constrain the close binary properties of early-B primaries in the MW (summarized in Table~6).  We now wish to compare these properties to spectroscopic observations of early-type stars in the MW.  This will demonstrate consistency between the eclipsing and spectroscopic methods of inferring the close binary parameters.  As with eclipsing systems, observations of spectroscopic binaries are biased toward systems with edge-on orientations and massive secondaries.  For each of the following spectroscopic samples, we must consider their sensitivity and completeness toward low-mass companions so that we can accurately compare ${\cal F}_{\rm close}$.

\begin{table}[H]
\small
\begin{tabular}{|c|c|c|c|c|c|c|}
\hline
  Spec. Type &      Method   & ${\cal F}_{\rm twin}$ & $\gamma_{\rm q}$ & $\gamma_{\rm P}$  & ${\cal F}_{\rm close}$  & Sample Reference \\
\hline
    Late-B   & \multirow{2}{*}{Spectroscopic} & \multirow{2}{*}{0.06\,$\pm$\,0.03}      &  \multirow{2}{*}{$-$1.2\,$\pm$\,0.4}  & \multirow{2}{*}{$-$0.3\,$\pm$\,0.4}    &  0.16\,$\pm$\,0.06 & \multirow{2}{*}{\citet{Levato1987}}  \\\cline{1-1}\cline{6-6}   
   Early-B   &               &                       &                     &                      &  0.22\,$\pm$\,0.07 &         \\
\hline
    Early-B  & Eclipsing     &    0.16\,$\pm$\,0.10   &   $-$0.9\,$\pm$\,0.8 &   $-$0.4\,$\pm$\,0.3~ &  0.21\,$\pm$\,0.05 & \citet{Lefevre2009} \\
\hline
    Early-B & Spectroscopic  & 0.06\,$\pm$\,0.05      &  $-$0.9\,$\pm$\,0.4  & ~0.2\,$\pm$\,0.5      &  0.23\,$\pm$\,0.06 & \citet{Abt1990} \\  
\hline
   O        & Spectroscopic  & 0.08\,$\pm$\,0.06      &  $-$0.2\,$\pm$\,0.5  & $-$0.5\,$\pm$\,0.3    &  0.31\,$\pm$\,0.07 & \citet{Sana2012}\\
\hline
\end{tabular}
\caption{Milky Way comparison of our fitted binary properties for early-type stars based on spectroscopic radial velocity observations to our analysis of Hipparcos MW eclipsing binaries. The close binary fraction, i.e. the fraction of systems which have a companion with orbital period $P$ = 2\,-\,20 days and mass ratio $q$ $>$ 0.1,  nearly doubles between late-B and O spectral type primaries. Other parameters are fairly consistent with a negligible excess twin fraction ${\cal F}_{\rm twin}$~$\approx$~7\%, a mass-ratio distribution weighted toward low-mass companions with $\gamma_{\rm q}$~$\approx$~$-$0.9, and a period distribution with $\gamma_{\rm P}$~$\approx$~$-$0.3 that is slightly skewed toward shorter periods relative to \"{O}pik's law.}
\end{table}

In the spectroscopic survey of  78 B-type stars in the Sco-Cen association, \citet{Levato1987} found 15 systems with $P$~=~2\,-\,20 days.  Their sample was complete to velocity semi-amplitudes of $K$~$\gtrsim$~15~km~s$^{-1}$.  Assuming a typical primary mass of $M_1$~$\approx$~5\,\Msun\ for a mid B-type star,  a representative inclination of $i$~$\approx$~50$^{\rm o}$, and their mean orbital period of $P$~$\approx$~6~days, then the corresponding sensitivity is coincidentally $q$~$\approx$~0.10. Since we do not need to correct for incompleteness down to $q$~=~0.1, the close binary fraction is ${\cal F}_{\rm close}$~=~15\,/\,78~=~(19\,$\pm$\,5)\%.  If we divide the sample into late-type ($\ge$\,B5) and early-type ($\le$\,B4) groups, then the close binary fractions would be ${\cal F}_{\rm close}$~=~(16\,$\pm$\,6)\% and  (22\,$\pm$\,7)\%, respectively.  

  Using these ${\cal N}$ = 15 systems in the \citet{Levato1987} catalog, we fit the orbital period distribution ${\cal U}_{\rm P}$ based on the theoretical parametrization in Eq. 1. To constrain $\gamma_{\rm P}$, we maximize the likelihood function L($\gamma_{\rm P}$) = $\prod_{k=1}^{\cal N} {\cal U}_{\rm P}(P_k | \gamma_{\rm P})$\,d(log\,$P$), where we ensure ${\cal U}_{\rm P}$ integrates to unity in this instance.  We repeat this procedure ${\cal N}$ times with  delete-one jackknife resamplings of the data to quantify the error.   We find $\gamma_{\rm P}$~=~$-$0.3\,$\pm$\,0.4, i.e. a distribution slightly skewed toward shorter periods but still consistent with \"{O}pik's law.  

  We also use these 15 systems to estimate a statistical mass-ratio distribution ${\cal U}_{\rm q}$.  For the three double-lined spectroscopic binaries with well-defined orbits, we determine $q$ simply from the ratio of the observed velocity semi-amplitudes.  For the remaining 12 systems, primarily single-lined spectroscopic binaries, we determine the primary mass $M_1$ from the spectral type, assume a random inclination in the interval $i$~=~10$^{\rm o}$\,-\,80$^{\rm o}$ for each system $k$, and then utilize the listed mass function $f$($M$) to estimate a statistical mass-ratio $q_k$. Using our parametrization in Eq. 2, we then maximize the likelihood function L($\gamma_{\rm q}$, ${\cal F}_{\rm twin}$) = $\prod_{k=1}^{\cal N} {\cal U}_{\rm q}(q_k | \gamma_{\rm q}, {\cal F}_{\rm twin})$\,d$q$, where we only include systems with statistical mass-ratios in the interval $q_k$ = 0.1\,-\,1.0.  To quantify the error, we repeat this process ${\cal N}$ times with delete-one jackknife resamplings of the data,  where we evaluate each of the systems without a dynamical mass ratio at a different random inclination.  We find a mass-ratio distribution weighted toward low-mass companions with $\gamma_{\rm q}$~=~$-$1.2\,$\pm$\,0.4, and a small excess twin fraction of ${\cal F}_{\rm twin}$~=~0.06\,$\pm$\,0.03.  We report these results in Table 6.  

In the magnitude-limited sample of early-B stars, \citet{Abt1990} corrected for classical Malmquist bias and found 16 out of 109 systems to be spectroscopic binaries with $P$~=~2\,-\,20~days. They were only sensitive down to velocity semi-amplitudes of $K$~$\gtrsim$~20~km~s$^{-1}$, but reported incompleteness factors down to $M_2$~$\approx$~0.7\,\Msun\ of $I$~$\approx$~1.4 for $P$~=~0.36\,-\,3.6~days and $I$~$\approx$~1.8 for $P$~=~3.6\,-\,36~days.  Given their nominal primary mass of $M_1$~$\approx$~8\,\Msun, we adopt an intermediate incompleteness factor of $I$\,=\,1.6 to correct down to $q$~$\approx$~0.1 for our systems of interest with $P$~=~2\,$\mbox{-}$\,20~days.  This results in a close binary fraction of ${\cal F}_{\rm close}$~=~16\,$\times$\,1.6\,/\,109~=~(23\,$\pm$\,6)\%, consistent with the early-B subsample result we derived from the \citet{Levato1987} data.  

 We determine the period distribution ${\cal U}_{\rm P}$ and mass-ratio distribution ${\cal U}_{\rm q}$ for the \citet{Abt1990} survey using two methods.  First, we fit the 16 observed systems  using the same procedure utilized above for the \citet{Levato1987} sample.  We find $\gamma_{\rm P}$~=~0.1\,$\pm$\,0.4,  $\gamma_{\rm q}$~=~$-$0.8\,$\pm$\,0.3, and ${\cal F}_{\rm twin}$~=~0.07\,$\pm$\,0.04. Second, we use the values in Table 6 of \citet{Abt1990}, which have been corrected for incompleteness.  They estimate there to be $\approx$\,5.7 systems with $P$~=~1.7\,-\,3.6~days, i.e. $\approx\,$17.5 systems per decade of period at log\,$P$~$\approx$~0.4, and $\approx$\,34.4 systems with $P$~=~3.6\,-\,36 days, i.e. 34.4 systems per decade of period at log\,$P$~$\approx$~1.1.  These two data points imply a slope of $\gamma_P$~=~0.3. We then utilize their four bins of secondary masses for the 40.1 systems with $P$~$<$~36 days.  Minimizing the $\chi^2$ statistic between the four bins of data and our two-parameter formalism ${\cal U}_{\rm q}$, we find $\gamma_{\rm q}$~$\approx$~$-$1.0 and ${\cal F}_{\rm twin}$~$=$~0.05.  We adopt the average of the two methods so that $\gamma_{\rm P}$~=~0.2\,$\pm$\,0.5, $\gamma_{\rm q}$~=~$-$0.9\,$\pm$\,0.4, and ${\cal F}_{\rm twin}$~=~0.06\,$\pm$\,0.05 (see Table 6).

Based on spectroscopic observations of 71 O-type stars in various open clusters, \citet{Sana2012} found 21 systems with orbital periods $P$~=~2\,-\,20~days.  After they corrected down to $q$~=~0.1, they estimated there to be only $\approx$\,1 additional system that escaped their detection in this period range.  This results in a close binary fraction of ${\cal F}_{\rm close}$~=~(31\,$\pm$\,7)\%, which is slightly higher than the B-type results. 

We fit the period and mass-ratio distributions for these 21 systems using the same method as for the \citet{Levato1987} sample.  We find $\gamma_P$~=~$-$0.5\,$\pm$\,0.3, which is consistent with their result of ${\cal U}_{\rm P}$~$\propto$~(log\,$P$)$^{-0.55\,\pm\,0.22}$\,d(log\,$P$) for all their spectroscopic binaries (note slightly different parametrization).  We also find $\gamma_{\rm q}$~$=$~$-$0.2\,$\pm$\,0.5 and ${\cal F}_{\rm twin}$~=~0.08\,$\pm$\,0.06, consistent with their fit of $\gamma_{\rm q}$~$=$~$\kappa$~$=$~$-$0.1\,$\pm$\,0.6 to all the systems in their sample. This result for the mass-ratio distribution is fairly robust because 18 of the 21 systems were double-lined spectroscopic binaries with dynamical mass ratios.  However, the formal error bar on the derived $\gamma_{\rm q}$ is quite large, so that the fit is still consistent with the lower values of $\gamma_q$  measured for the previous populations.   
 
We compare the close binary parameters for the three spectroscopic samples and the Hipparcos eclipsing binary sample in Table 6.  The only clear trend is an increasing close binary fraction with primary mass so that ${\cal F}_{\rm close}$ nearly doubles between late-B and O type stars.  Assigning $\langle M_1 \rangle$~=~4\,\Msun, 10\,\Msun, and 25\,\Msun\ to late-B, early-B, and O spectral types, respectively, the Pearson correlation coefficient of log\,$M_1$ versus log\,${\cal F}_{\rm close}$ for the five data points in Table 6 is $r$~=~0.99. This highly significant correlation implies that $M_1$ and ${\cal F}_{\rm close}$ are related via a simple power-law, which we find to be ${\cal F}_{\rm close}$ = 0.22($M_1$/10\Msun)$^{0.4}$.   All of the populations are consistent with a small twin fraction ${\cal F}_{\rm twin}$~$\approx$~7\%, mass-ratio distribution that favors low-mass companions with $\gamma_{\rm q}$~$\approx$~$-$0.9, and a period distribution with $\gamma_{\rm P}$~$\approx$~$-$0.3 that is skewed toward shorter periods compared to \"{O}pik's law.  The fact that all the derived binary properties derived from the eclipsing and spectroscopic binary samples are in agreement is testament to the robustness of our eclipsing binary models and the validity of ${\cal F}_{\rm close}$ reported for the different environments in \S3.  

\section{Discussion}

\subsection{Summary}

We have analyzed four different samples of eclipsing binaries with early-B primaries: one in the MW with $\langle$log($Z$/\Zsun)$\rangle$ = 0.0, two in the LMC with $\langle$log($Z$/\Zsun)$\rangle$ = $-$0.4, and one in the SMC with $\langle$log($Z$/\Zsun)$\rangle$ = $-$0.7. The fractions of early-B stars which exhibit deep eclipses 0.25~$<$~$\Delta$m\,(mag)~$<$~0.65 with orbital periods 2~$<$~$P$\,(days)~$<$~20 span a narrow range of ${\cal F}_{\rm deep}$ = (0.7\,-\,1.0)\% among all four populations (Table 1).  The OGLE-II LMC and SMC observations become incomplete toward shallower eclipses, while the OGLE-III LMC and Hipparcos MW observations are complete to $\Delta$m = 0.1.  For these latter two surveys, ${\cal F}_{\rm med}$ = 1.9\% of early-B stars exhibit eclipses 0.1 $<$ $\Delta$m $<$ 0.65 with $P$ = 2\,-\,20 days (Table 1).  The consistency of these results are model independent, demonstrating that the eclipsing binary fractions do not vary with metallicity.  

All four samples have similar eclipse depth distributions ${\cal O}_{\Delta {\rm m}}$ across the intervals over which their respective surveys are complete (Figure 1).  Based on the larger and more complete OGLE-III LMC sample, we find a simple power-law fit ${\cal S}_{\Delta {\rm m}}$\,d($\Delta$m) $\propto$ ($\Delta$m)$^{-1.65\,\pm\,0.07}$\,d($\Delta$m), which is significantly steeper than the distribution ${\cal S}_{\Delta {\rm m}}$\,d($\Delta$m) $\propto$ ($\Delta$m)$^{-1.0}$\,d($\Delta$m) we would expect if the companions were selected from a uniform mass-ratio distribution.   All four samples also have observed period distributions ${\cal O}_{\rm deep}(P)$ or ${\cal O}_{\rm med}(P)$ that are slightly skewed toward shorter periods relative to \"{O}pik's  prediction of ${\cal S}_{\rm deep}(P)$\,d(log\,$P$) $\propto$ ${\cal S}_{\rm med}(P)$\,d(log\,$P$) $\propto$ $P^{-2/3}$\,d(log\,$P$) (Figure 2).  The OGLE-II SMC distribution is especially weighted toward shorter periods, but this sample may be slightly incomplete for modest eclipse depths $\Delta$m = 0.25\,-\,0.30 mag and longer orbital periods $P$ = 10\,-\,20 days.  It would be worthwhile to examine this feature once an OGLE-III SMC eclipsing binary catalog becomes available.

In order to correct for geometrical selection effects and incompleteness toward low-mass companions, we employed detailed \textsc{nightfall} light curve models and performed thousands of Monte Carlo simulations for various binary populations.    By minimizing the $\chi^2$ statistics between the observed distributions ${\cal O}$ and our models ${\cal M}$, we were able to constrain the underlying properties ${\cal U}$ of the close binaries in each of our samples.  In our models, we considered a multitude of systematic effects including tidal distortions, mutual irradiation, limb darkening, stellar evolution and Roche lobe filling, third light contamination due to stellar blending and triple star systems, eccentric orbits, uncertainties in dust extinction,  and Malmquist bias. 

  The four fitted model parameters $\gamma_{\rm q}$, ${\cal F}_{\rm twin}$, $\gamma_{\rm P}$, and ${\cal F}_{\rm close}$ for all four eclipsing binary samples are fairly consistent with each other.  The mean mass-ratio exponents span  $\gamma_{\rm q}$~= $-$1.0\,-\,$-$0.6 for the four samples (Table~2 and Figure 7), suggesting the mass-ratio distribution ${\cal U}_{\rm q}$ $\appropto$ $q^{\gamma_{\rm q}}$\,d$q$ is weighted toward lower mass companions relative to a uniform distribution with $\gamma_{\rm q}$ = 0. An excess of twins with $q$ $>$ 0.9 comprise a small fraction ${\cal F}_{\rm twin}$~=~(4\,-\,16)\%  of all companions with $q$~$>$~0.1 (Table~2 and Figure 7).  The period distributions are slightly skewed toward shorter periods relative to \"{O}pik's law, giving $\gamma_{\rm P}$ = $-$0.9\,-\,$-$0.1  in the relation ${\cal U}_{\rm P}$ $\propto$ $P^{\gamma_{\rm P}}$\,d(log\,$P$) (Table 3 and Figures~10~\&~12).  Finally, the close binary fractions with $q$ $>$ 0.1 and $P$ = 2\,-\,20 days span a narrow range of ${\cal F}_{\rm close}$ = (16\,-\,25)\% (Table 5 and Figure 12).  None of these parameters exhibited a trend with metallicity, signifying that the close  binary properties do not vary with metallicity across the interval $-$0.7~$<$~log($Z$/\Zsun)~$<$~0.0.   

We emphasize that these model parameters are only valid for $q$ $>$ 0.1 and $P$ = 2\,-\,20 days, and should not be extrapolated toward lower mass companions or longer orbital periods.  Moreover, these quantities represent the mean values in our parameter space because we have assumed the mass-ratio distribution ${\cal U}_{\rm q}$ is independent of the orbital period $P$.  The large OGLE-III LMC medium eclipse depth sample exhibits a statistically significant trend between $P$ and $\Delta$m, and we will investigate this feature in more detail in a future study.   Nevertheless, all four samples of eclipsing binaries exhibited weak or no correlations between $P$ and $\Delta$m with Spearman rank coefficients $|\rho|$ $<$ 0.15. In addition, the probabilities of observing medium eclipses ${\cal P}_{\rm med}$($P$) are relatively independent of the underlying mass-ratio distribution ${\cal U}_{\rm q}$ (see \S3.3.2). The close binary fraction ${\cal F}_{\rm close}$ for the OGLE-III LMC population will therefore not vary beyond the cited errors, even when we consider a period-dependent mass-ratio distribution.

\subsection{Comparison with Previous Studies}

In \S4, we examined three samples of spectroscopic binaries in the MW with early-type primaries \citep{Levato1987, Abt1990, Sana2012}.   These observations demonstrated that the close binary fraction increased by nearly a factor of two between late-B type primaries with ${\cal F}_{\rm close}$ $\approx$ 16\% and O-type primaries with ${\cal F}_{\rm close}$ $\approx$ 31\%.  The three samples were consistent with a negligible excess twin fraction ${\cal F}_{\rm twin}$~$\approx$~7\%, a mass-ratio distribution weighted toward low-mass companions with $\gamma_{\rm q}$~$\approx$~$-$0.9, and a period distribution with $\gamma_{\rm P}$~$\approx$~$-$0.3 that is slightly skewed toward shorter periods relative to \"{O}pik's law.  The only outlier beyond the 1$\sigma$ level was the overall mass-ratio distribution of the \citet{Sana2012} sample, which we fitted to have $\gamma_{\rm q}$~=~$-$0.2\,$\pm$\,0.5.  More recently, however, \citet{Sana2013} found a lower value and tighter constraint of $\gamma_q$~$=$~$\kappa$~$=$~$-$1.0\,$\pm$\,0.4 based on spectroscopic observations of O-type stars in 30 Doradus, which is even more consistent with our mean value.  The fact that the close binary fractions and properties inferred from spectroscopic binaries match the parameters derived from our eclipsing binary samples is testament to the robustness of our models.

 There may indeed be a narrow peak of twins in the mass-ratio distribution so that ${\cal U}_{\rm q} (q \approx 1)$ is several times the value of ${\cal U}_{\rm q} (q \approx 0.8)$.  However, this twin contribution represents a small fraction of the total population of secondaries in the entire interval 0.1 $<$ $q$ $<$ 1.  Based on a sample of 21 detached eclipsing binaries in the SMC with massive primaries, $P$ $<$ 5 days, and well-determined spectroscopic orbits, \citet{Pinsonneault2006} estimated a modest excess twin fraction of ${\cal F}_{\rm twin}$ =  20\,-\,25\%.  However, they assumed their underlying uniform mass-ratio distribution could be extrapolated below their detection limit of $q$ $\approx$ 0.55, so they expected relatively few systems below their survey sensitivity.  If instead the low-$q$ tail was replaced with our fitted mean value of $\gamma_{\rm q}$ = $-$1.0\,-\,$-$0.6, depending on the sample, then the twin fraction would be reduced to ${\cal F}_{\rm twin}$ = (5\,-\,10)\%, which is consistent with our results.  Because we find the overall mass-ratio distribution to be weighted toward lower masses with $\gamma_{\rm q}$ $\approx$ $-$0.9, the relative contribution of twin systems with $q$ $\gtrsim$ 0.9 is small compared to all secondaries across the interval 0.1 $<$ $q$ $<$ 1.

\citet{Mazeh2006} used OGLE-II LMC eclipsing binary data to derive a close binary fraction of 0.7\%.  Our value of ${\cal F}_{\rm close}$ = (16\,$\pm$\,6)\% for this population is a factor of $\approx$20 higher for four reasons.  First, \citet{Mazeh2006} only included systems with  orbital periods $P$ = 2\,-\,10 days while we extended our sample to include orbital periods up to $P$ = 20 days.  Assuming \"{O}pik's law, we would expect our close binary fraction to be 40\% higher, a minor contribution to the overall discrepancy.  Second, our samples contained early-B primaries with $-$3.8~$<$~M$_{\rm I}$~$<$~$-$1.3 while \citet{Mazeh2006} considered late-B stars with $-$1.8~$<$~M$_{\rm I}$~$<$~0.2.  The close binary fraction rapidly increases with primary mass (see \S4), so that ${\cal F}_{\rm close}$ for early-B stars is $\approx$1.5 times the late-B value.  Third, although \citet{Mazeh2006} accounted for geometrical selection effects, they did not correct for incompleteness toward small, low-mass secondaries.   The increase in the eclipsing binary fraction from ${\cal F}_{\rm deep}$ = 0.7\% to  ${\cal F}_{\rm med}$ = 1.9\% already suggests that the increased sensitivity of the OGLE-III survey could find three times more eclipsing systems.  In \S3.3.2, we showed that correcting for mass-ratio incompleteness {\it alone} increased the inferred close binary fraction by a factor of ${\cal C}_{\rm deep,q}$ $\approx$ 3.   Finally,  our reduction in ${\cal F}_{\rm close}$ due to Malmquist bias of the second kind by a factor of ${\cal C}_{\rm Malm}$ = 0.73 is a not as severe as the factor of ${\cal C}_{\rm Malm}$ = 0.48 implemented by \citet{Mazeh2006}.  This is partially because the average luminosity of the eclipsing companions was fainter than the $\langle\Delta$M$_{\rm I} \rangle$ = 0.5 mag I-band excess assumed by \citet{Mazeh2006}, but also because we accounted for other binaries with luminous secondaries outside our eclipsing binary parameter space of eclipse depths and orbital periods.

\subsection{Conclusions}

Weighting our four samples of eclipsing binaries and the three samples of spectroscopic binaries, we find the best overall model parameters to be ${\cal F}_{\rm twin}$ = 0.07\,$\pm$\,0.05, $\gamma_{\rm q}$ = $-$0.9\,$\pm$\,0.3, and $\gamma_{\rm P}$ = $-$0.3\,$\pm$\,0.3.  The close binary fraction increases with primary mass according to ${\cal F}_{\rm close}$ = (0.22\,$\pm$\,0.05)($M_1$\,/\,10\Msun)$^{0.4}$.  None of these properties exhibited statistically significant trends with metallicity across the interval $-$0.7~$<$~log($Z$/\Zsun)~$<$~0.0, demonstrating the close binary properties of massive stars are fairly independent of metallicity.    Any observed variations in the rates or properties of massive star or binary evolution within this metallicity range must derive from metallicity-dependent stellar physical processes, and not on the initial conditions of the MS binaries themselves.

We acknowledge support from NSF grant AST-1211843.  M.\,M. thanks Tsevi Mazeh, Ian Czekala, and Tanmoy Laskar for enlightening discussions of eclipsing binaries and statistics.

\bibliographystyle{apj}                       
\bibliography{bibliography}

\end{document}